\documentclass[traditabstract]{aa}
\usepackage{color, colortbl}
\usepackage{graphicx}
\usepackage{txfonts}
\usepackage{pstricks}
\usepackage{natbib}
\usepackage{multirow}
\bibpunct{(}{)}{;}{a}{}{,}

\def\0{\phantom0}

\begin{document}

\title{Interpolating point spread function anisotropy}

\author{M.~Gentile\inst{1}, F.~Courbin\inst{1} \and G.~Meylan\inst{1}}

\institute{Laboratoire d'astrophysique, Ecole Polytechnique F\'ed\'erale de Lausanne (EPFL), Observatoire de Sauverny, CH-1290 Versoix, Switzerland, \email{marc.gentile@epfl.ch} \label{inst1}
}

\date{Received; accepted }
 
\abstract{Planned wide-field weak lensing surveys are expected to reduce the statistical errors on the shear field to unprecedented levels. In contrast, \mbox{systematic} errors like those induced by the convolution with the point spread function (PSF) will not  benefit from that scaling effect and will require very accurate modeling and correction. While numerous methods have been devised to carry out the PSF correction itself, modeling of the PSF shape and its spatial variations across the instrument field of view has, so far, attracted much less attention. This step is nevertheless crucial because the PSF is only known  at star positions while the correction has to be performed at any position on the sky. A reliable interpolation scheme is therefore mandatory and a popular approach has been to use low-order bivariate \mbox{polynomials}. In the present paper, we evaluate four other classical spatial interpolation methods based on splines (B-splines), inverse distance weighting (IDW), radial basis functions (RBF) and ordinary Kriging (OK). These methods are tested on the Star-challenge part of the GRavitational lEnsing Accuracy Testing 2010 (GREAT10) simulated data and are compared with the classical polynomial fitting (Polyfit). In all our methods we model the PSF using a single Moffat profile and we interpolate the fitted parameters at a set of required positions. This allowed us to win the Star-challenge of GREAT10, with the B-splines method. However, we also test all our interpolation methods independently of the way the PSF is modeled, by interpolating the GREAT10 star fields themselves (i.e., the PSF parameters are known exactly at star positions). We find in that case RBF to be the clear winner, closely followed by the other local methods, IDW and OK. The global methods, Polyfit and B-splines, are largely behind, especially in fields with (ground-based) turbulent PSFs. In fields with non-turbulent PSFs, all interpolators reach a variance on PSF systematics $\sigma_{sys}^2$ better than the $1\times10^{-7}$ upper bound expected by future space-based surveys, with the local interpolators performing better than the global ones. 

} 

\titlerunning{Point Spread Function interpolation}
%%% Titre pour version referee
\maketitle
%
%________________________________________________________________

\section{Introduction}
\label{intro}

The convolution of galaxy images with a Point Spread Function (PSF) is among the primary sources of systematic error in weak lensing measurement. The isotropic part of the PSF kernel makes the galaxy shape appear rounder, while the anisotropic part introduces an artificial shear effect that may be confused with the genuine shear lensing signal.

To tackle these issues, various PSF correction methods have been proposed \citep{KSB1995a, LuppinoKaiser1997, Hoekstra1998, Kaiser2000a, BernsteinJarvis2002, HirataSeljak2003, RefregierBacon2003} and some of them implemented as part of shear measurement pipelines \citep{HeymansSTEPI2006, MasseySTEP22007, BridleGREAT082010}.
However, these correction  schemes do not have built-in solutions for addressing another problem: the spatial variation of the PSF across the instrument field of view that may arise, for instance, from imperfect telescope guidance, optical aberrations or atmospheric distortions.

A non-constant PSF field implies the PSF is no longer accurately known at galaxy positions and must then be estimated for the accurate shape measurement of galaxies. Bivariate polynomials, typically used as interpolators for this purpose, have in several cases been found unable to reproduce sparse, multi-scale or quickly varying PSF anisotropy patterns \citep{Hoekstra2004, JarvisJain2004, VanWaerbeke2002, VanWaerbekeMellierHoekstra2005, JeeTyson2011}.\newline\indent
This raises the question of whether there exists alternative PSF models and interpolation schemes better suited for PSF estimation than those used so far. Indeed, it seems important to improve this particular aspect of PSF modeling in the perspective of future space-based missions such as Euclid or advanced ground-based telescopes like the LSST \citep{JeeTyson2011}.

Only recently has the PSF variation problem begun to be taken seriously with, notably, the advent of the GRavitational lEnsing Accuracy Testing 2010 (GREAT10) Star Challenge, one of the two GREAT10 challenges \citep{GREAT10Handbook2010, GREAT10StarChallengeResults2012}. The Star Challenge images have been designed to simulate a variety of typical position-varying PSF anisotropy patterns and competing PSF interpolation methods were judged on their ability to reconstruct the true PSF field at asked, non-star positions.

The Star Challenge gave us the opportunity to evaluate a number of alternative schemes suitable for the interpolation of realistic, spatially-varying PSF fields. The objective of this paper is twofold: (1) to describe our approach for tackling the problems raised by the Star Challenge and to discuss our results; (2) to perform a comparative analysis of the different interpolation methods after applying them on the Star Challenge simulations.

Our paper is thus organized as follows. We begin by reviewing the most commonly used PSF representation and interpolation schemes in Sect.~\ref{section:existing PSF interpolation schemes} and continue with a overview of the interpolation schemes mentioned above in Sect.~\ref{section:PSF interpolation schemes}. We then describe our PSF estimation pipeline and analyze our results in Sects.~\ref{section:applying on GREAT10 data} and \ref{section:results analysis}. Lastly, in Sect.~\ref{section:comparative analysis}, we measure the respective accuracy of all methods based on the solutions made available after completion of the challenge and discuss the merits of each method. We conclude in Sect.~\ref{section:conclusions}.

\section{An overview of existing PSF interpolation schemes}
\label{section:existing PSF interpolation schemes}

\begin{figure}[b]
%\scalebox{0.30}{\includegraphics{fig_local_weighted_interpolation.pdf}} 
%\resizebox{\hsize}{!}{\scalebox{0.30}{\includegraphics{fig_local_weighted_interpolation.pdf}} }
\resizebox{\hsize}{!}{\includegraphics[trim=5mm 5mm 5mm 5mm, clip]{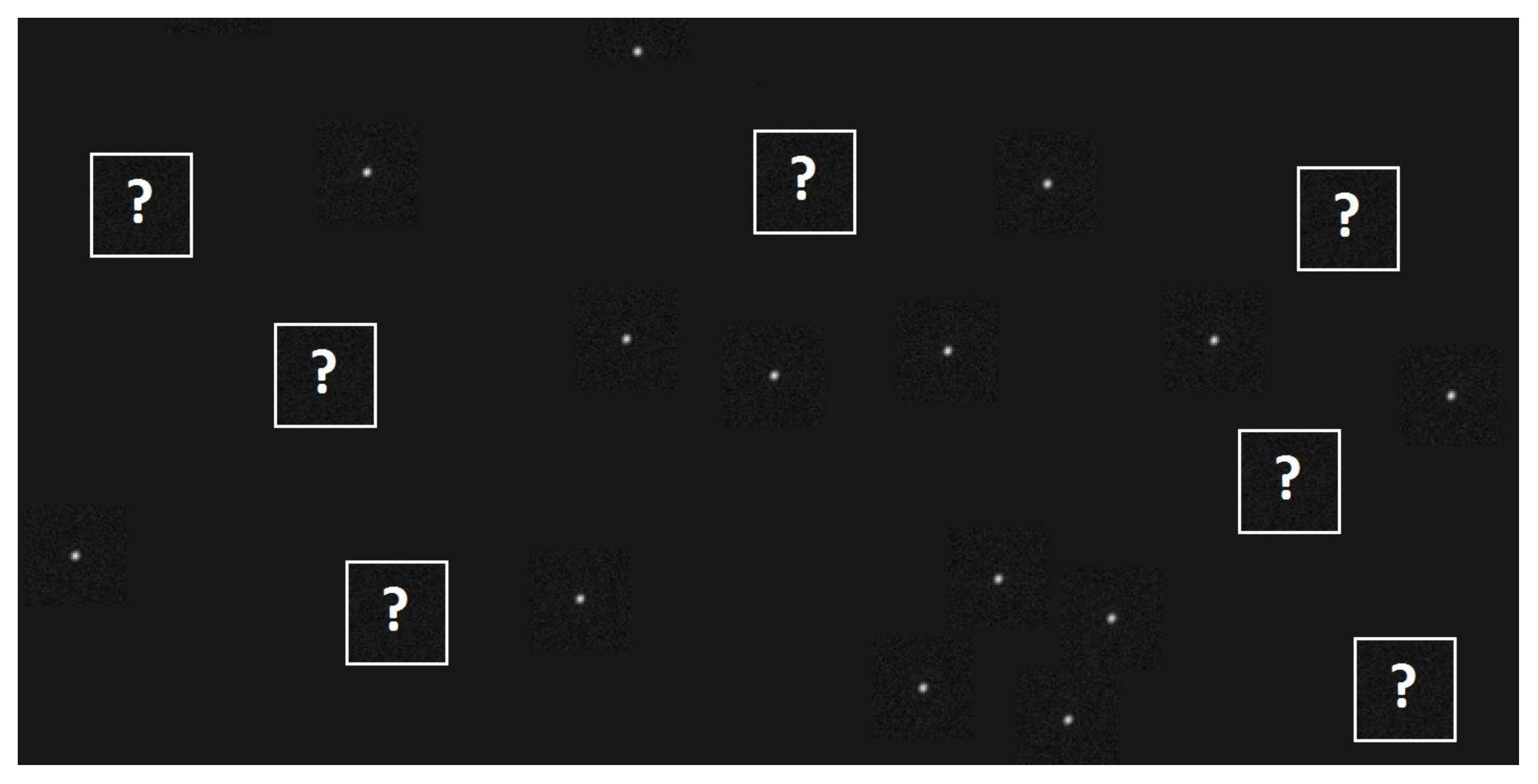}}
\caption{Interpolating a spatially-varying PSF field. The illustrated field is a subset of an actual GREAT10 Star Challenge PSF field.}
\label{fig:PSF interpolation}
\end{figure}

Before correcting galaxies in the image for a spatially-varying PSF field, every shear measurement pipeline has, in one way or another, to interpolate the PSF between the stars, as illustrated in Fig.~\ref{fig:PSF interpolation}. The way this is best achieved depends essentially on the PSF model used and on the PSF interpolation algorithm. The PSF model defines which features of the PSF are to be represented, which also determines on which quantities spatial interpolation is performed. The role of the interpolation scheme, on the other hand, is to apply a prediction algorithm to find the best estimates for those quantities.

In the KSB method \citet{KSB1995a} and its KSB+ variant \citep{LuppinoKaiser1997, Hoekstra1998}, the relevant features of the PSF model are its ellipticity and size, which are estimated from the second-order geometrical moments of the PSF image. The main idea behind the PSF correction scheme is that the PSF distortion on a galaxy image can be well described by a small but highly anisotropic kernel $q$, convolved with a large, circular seeing disk. To find the appropriate $q$ for galaxies, the values of $q^{*}$ at star positions (and sometimes the so-called "smear" and "shear" polarization tensors $P^{sm*}$ and $P^{sh*}$) are interpolated across the image. For doing so, the typical procedure is to fit a second or third-order bivariate polynomial function. 

Exactly which quantity is interpolated and which order is used for the polynomial depends on the KSB+ implementations. See e.g. \citet{HeymansSTEPI2006}, Appendix~\ref{section:appendix PSF prediction pipeline}, \citet{MasseySTEP22007} and recently published studies using KSB+ \citep{Hoekstra1998, CloweSchneider2002, Heymans2005, Hetterscheidt2007, Paulin-Henriksson2007, fu2008, Umetsu2010}.\\

A model representing a PSF as only a size and first-order deviation from circularity certainly appears quite restrictive. One can instead look for an extensive, but compact description of the PSF image, better suited to operations like noise filtering or deconvolution. A natural approach is to characterize the full PSF as a compact, complete set of orthogonal basis functions provided in analytical form, each basis being associated with a particular feature of the image (shape, frequency range, etc.). Ideally, this would not only simplify galaxy deconvolution from the PSF but also allow to better model the spatial variation of the PSF across the field of view.

\citet{BernsteinJarvis2002} and \citet{Refregier2003, RefregierBacon2003, MasseyRefregier2005} have proposed PSF expansions based on the eigenfunctions of the two-dimensional quantum harmonic oscillator, expressed in terms of Gauss-Laguerre orthogonal polynomials \citep{AbramowitzStegun1965}. These functions can be interpreted as perturbations around a circular or elliptical Gaussian. The effect of a given operation (such as shear or convolution), on an image can then be traced through its contribution on each coefficient in the basis function expansion. For instance, the second-order $f_{2,2}$ coefficient of a Shapelet is the ellipticity estimator based on the Gaussian-weighted quadrupole moments used in KSB.

Modeling the PSF variation patterns with Shapelets typically involves the following steps: stars are expanded in terms of Shapelet basis functions and the expansion coefficients for each of the basis functions are fitted with a third or fourth-order polynomial. The interpolated values of the Shape let coefficients are then used to reconstruct the PSF at galaxy positions.
 
This scheme has been successfully applied to several weak lensing cluster studies \citep{JeeWhiteBenitez2005, JeeWhiteford2005, JeeWhiteFord2006, JeeFordIllingworth2007, Berge2008, Romano2010}. However, it has been argued \citep{JeeBlakeslee2007, Melchior2010} that even a high-order Shapelet-based PSF model is unable to reproduce extended PSF features (such as its wings) and that the flexibility of the model makes it vulnerable to pixelation and noise. So, although the level of residual errors after Shapelets decomposition appears low enough for cluster analysis, it may prove too high for precision cosmic shear measurement.\\

Actually, it is not clear if there exists any set of basis functions expressed in analytical form that is capable of accurately describing all the signal frequencies contained in the PSF. An alternative approach is to decompose the PSF in terms of \mbox{basis} functions directly derived from the data through Principal Component Analysis (PCA), as pioneered by \citet{Lauer2002}, \citet{Lupton2001}. This approach is supposed to yield a set of basis function, the so-called "Principal Components", optimized for a particular data configuration and sorted according to how much they contribute to the description of the data.

In practice, two main procedures have been experimented that essentially depend on the type data where PCA is applied.  
\citet{JarvisJain2004} and \citet{Schrabback2010} fit selected components of the PSF (e.g. ellipticity or KSB anisotropy kernel) across all image exposures with a two-dimensional polynomial of order 3 or 4. PCA analysis is performed on the coefficients of the polynomial, which allows the large-scale variations of the PSF in each exposure to be expressed as a weighted sum of a small set of principal components. A further, higher-order polynomial fit is then conducted on each exposure to capture more detailed features of the PSF.

On the other hand, and more recently, \citet{JeeBlakeslee2007}, \citet{NakajimaBernstein2010} and \citet{JeeTyson2011} experimented a different procedure for modeling the variation of the Hubble Space telescope (HST) ACS camera and a simulated Large Synoptic Survey Telescope (LSST) PSF. Instead of applying PCA on polynomial coefficients, they perform a PCA decomposition on the star images themselves into a basis made of the most discriminating star building blocks. Each star can then be expanded in terms of these "eigenPSFs" and the spatial variation of their coefficients in that basis is modeled with a bivariate polynomial.

Regardless of the procedure used, the PCA scheme proves superior to wavelets and Shapelet for reproducing smaller-scale features in the PSF variation pattern, thanks to improved PSF modeling and the use of higher-order polynomials. In the case of \citet{JarvisJain2004}, applying PCA across all exposures allowed to compensate for the small number of stars available per exposure. Moreover, PCA is not tied to any specific PSF model.

It should be noted, however, that at least two factors may limit the performance of PCA in practical weak lensing applications: the first is that the PCA algorithm is only able to capture linear relationships in the data and thus may fail to reproduce some types of high-frequency variation patterns; the other is that PCA misses the components of the PSF pattern that are random and uncorrelated, such as those arising from atmospheric turbulence. How serious these limitations prove to be and how they can be overcome need to be investigated further \citep[e.g.][]{JainJarvisBernstein2006, Schrabback2010}.\\

All the above methods attempt to model PSF variation patterns in an empirical way by the application of some mathematical formalism. It may, on the contrary be more beneficial to understand which physical phenomena determine the structure of the PSF patterns and, once done, seek appropriates models for reproducing them \citep{JeeBlakeslee2007, Stabenau2007, JeeTyson2011}. The PSF of the HST ACS camera, for instance, has been studied extensively and in some cases, the physical origin of some of the patterns clearly identified. \citet{JeeBlakeslee2007} and \citet{JeeTyson2011} could relate the primary principal component to changes in telescope focus causes by constraints on the secondary mirror supporting structure and the "thermal breathing" of the telescope.

In fact, various combined effects make the PSF vary spatially or over time. Some patterns are linked to the behavior of the optical system of the telescope or the detectors. Others are related to mechanical or thermal effects that make the telescope move slightly during an observation. For ground-based instruments, refraction in the atmosphere and turbulence induce further PSF distortion.
 
Incorporating such a wide diversity of effects into a PSF variation model is not an easy task. However, according to \citet{Jarvis2008}, models of low-order optical aberrations such as focus and astigmatism can reproduce 90\% of the PSF anisotropy patterns found in real observation data. If so, physically-motivated models could provide an alternative or a complement to purely empirical methods such as PCA.

\section{Looking for better PSF interpolation schemes}
\label{section:PSF interpolation schemes}
The analysis of commonly used PSF interpolation schemes in the previous section has shown that the range of PSF interpolation algorithms is actually quite restricted: almost always the quantities used to characterize the PSF are fitted using a bivariate polynomial function.

But it is important to acknowledge there may exist alternative interpolation schemes that would prove more effective for that purpose than polynomial fitting.  Beyond this, it is essential to recognize the goal here is not to only interpolate changes in the PSF but also to perform a \emph{spatial} interpolation of such changes.

Interpolation \citep[e.g.][]{NumericalRecipes2007} is commonly understood as the process of estimating of values at location where no sample is available, based on values measured at sample locations. \emph{Spatial interpolation} differs from regular interpolation in that it can take into account and potentially exploit spatial relationships in the data. In particular, it is often the case that points close together in space are more likely to be similar than points further apart. In other words, points may be spatially autocorrelated, at least locally. Most spatial interpolation methods attempt to make use of such information to improve their estimates.

After a critical review of polynomial fitting, we consider and discuss alternative spatial interpolation schemes for modeling PSF variation patterns. 

\subsection{A critical view of polynomial fitting}
\label{subsection:polynomial fitting}

In the context of spatial interpolation, fitting polynomial functions of the spatial coordinate $\mathbf{x}={(x_i,y_i)}$ to the sampled $z(\mathbf{x})$ values of interest by ordinary least squares regression (OLS) is known as "Trend Surface Analysis" (TSA). The fitting process thus consists in minimizing the sum of squares for $(\hat z(\mathbf{x})-z(\mathbf{x}))$, assuming the data can be modeled as a surface of the form \begin{equation}\label{eq:polynomial surface}\hat z(\mathbf{x})=\sum_{r+s\leq{p}}{b_{rs}\, x^r\, y^s}\end{equation}The integer $p$ is the order of the trend surface (and the order of the underlying polynomial). Finding the $b_i$ coefficients is a standard problem in multiple regression and can be computed with standard statistical packages.

In the literature reviewed from the previous section, authors often justify their choice of polynomial fitting by arguing the PSF varies in a smooth manner over the image. Indeed trend surfaces are well suited to modeling broad features in the data with a smooth polynomial surface, commonly of order 2 or 3.

However, PSF variation patterns result from a variety of physical effects and even though polynomials may adequately reproduce the smoothest variations, there may exist several other types of patterns that a low-order polynomial function cannot capture. Polynomials are also quite poor at handling discontinuities or abrupt changes in the data. This concerns particularly sharp discontinuities across chip gaps and rapid changes often found near the corners of the image.

An illustrative example of the shortcomings just described was the detection of a suspicious non-zero B-mode cosmic shear signal in the VIRMOS-DESCART survey \citep{VanWaerbeke2001, VanWaerbeke2002}. After investigation \citep{Hoekstra2004, VanWaerbekeMellierHoekstra2005}, the small scale component of the signal was traced to the PSF interpolation scheme: the second-order polynomial function was unable to reproduce the rapid change in PSF anisotropy at the edges of the images. In fact, one of the main limitation of polynomials when used for interpolating PSF images in weak lensing studies lie in their inability to reproduce variations on scales smaller than the typical inter-stellar distance on the plane of the sky (often $\lesssim$ 1 arcmin at high galactic latitude).

Unfortunately there are no satisfactory solutions to these shortcomings. Increasing the order of the polynomial function does not help as it leads to oscillations while attempting to capture smaller-scale or rapidly-varying features. The $z(\mathbf{x})$ values may reach extremely (and unnaturally) small or large values near the edge or just outside the area covered by the data. Such extreme values can also create problems in calculations.

One way to alleviate such problems is to pack more densely the interpolating points closer to the boundaries, but this may not be easy to achieve in practice. \citet{Hoekstra2004} and \citet{VanWaerbekeMellierHoekstra2005} also obtained good results with an interpolator made of a polynomial function to model large-scale changes combined with a rational function to deal with small-scale variations. It is not clear, however, if this scheme can be safely applied on different data and this may require a significant amount of fine tuning.

In addition to the issues just mentioned, local effects in one part of the image may influence the fit of the whole regression surface, which makes trend surfaces very sensitive to outliers, clustering effects or observable errors in the $z(\mathbf{x})$. Finally, OLS regression implicitly assumes the $z(\mathbf{x})$ are normally distributed with uncorrelated residuals. These assumptions do not hold true when spatial dependence exists in the data.

Actually, the fact that trend surfaces tend to ignore any spatial correlation or small-scale variations can turn into an advantage to remove broad features of the data prior to using some other, finer-grained interpolator. Indeed, we saw in section \S1 that \citet{JarvisJain2004} took advantage of this feature in their PCA-based interpolation scheme. 

Most of the aforementioned limitations are rooted in the use of standard polynomials. One possible way out is to abandon trend surfaces altogether and use piecewise polynomials instead (especially Chebyshev polynomials), splines \citep{Deboor1978, Dierckx1995, Schumaker2007, Prenter2008} or alternative schemes that do not involve polynomials. Table~\ref{table:polynomial fitting} recalls the main advantages and disadvantages of polynomial fitting.

\begin{table}
\renewcommand{\arraystretch}{1.25}
\caption{Least squares polynomial fitting / Trend surface: Pros and cons}              % title of Table
\label{table:polynomial fitting}      % is used to refer this table in the text
%\centering                                      % used for centering table
\begin{tabular}{l p{7.5cm}}          % centered columns (4 columns)
\hline\hline                        % inserts double horizontal lines
\multicolumn{2}{l}{Least squares polynomial Fitting} \\    % table heading
\hline                                   % inserts single horizontal line
\multirow{2}{*}{Pros} & Simple and intuitive \\ 
                      & Fast to compute \\ 
\hline 
\multirow{6}{*}{Cons} & Usually only able to capture broad features (underfitting)  \\
		   & Increasing the order of polynomials does not improve and generally degrades accuracy (overfitting) \\
		   & High-order polynomials generate numerical issues (rounding errors, overflow, etc.) \\
                      & High sensitivity to outliers and fitting errors \\ 
                      & Local changes propagate to the whole polynomial surface \\ 
                      & No estimation of interpolation errors (deterministic) \\ 

\hline                                             %inserts single line
\end{tabular}
\end{table}

\subsection{Toward alternative PSF interpolation methods}
\label{subsection:alternative PSF interpolation schemes}
Having pointed out some important shortcomings of polynomial regression, it seems legitimate to look for alternative classes of interpolators. It is however, probably illusory to look for an ideal interpolation scheme that can describe equally well any kind of PSF variation structure. For instance the patterns of variation in a turbulent PSF are very different from those found in a diffraction-limited PSF. It is therefore probably more useful to identify which class of interpolators should be preferably used for a particular type of PSF pattern.

It is also key to realize that one does not need to reconstruct the entire PSF field: one only has to infer the PSF at specific galaxy positions based on its knowledge at sample star positions. This implies that the class of interpolation schemes applicable to the PSF variation problem is not restricted to surface fitting algorithms such as polynomial fitting, but also encompasses interpolation algorithms acting on scattered data.

Such data may also be considered as a partial realization of some stochastic process. In such case, it becomes possible to quantify the uncertainty associated with interpolated values and the corresponding interpolation method is referred to as a method for \emph{spatial prediction}. In this article we will neglect this distinction and use the generic term "spatial interpolation".

In fact, there are quite a few interpolation schemes that can be applied to model PSF changes. Over the years a large number of interpolation methods have been developed in many disciplines and with various objectives in mind. Spatial interpolators are usually classified according to their range (\emph{local} versus \emph{global}), the amount of smoothing (\emph{exact} versus \emph{approximate}) and whether they consider the data as a realization of some random process (\emph{stochastic} versus \emph{deterministic}).

%\begin{figure}[b]
%%\scalebox{0.30}{\includegraphics{fig_local_weighted_interpolation.pdf}} 
%%\resizebox{\hsize}{!}{\scalebox{0.30}{\includegraphics{fig_local_weighted_interpolation.pdf}} }
%\resizebox{\hsize}{!}{\includegraphics[trim=10mm 10mm 10mm 10mm, clip]{fig_local_neighborhood.pdf}}
%\caption{Local interpolation between a set of nearby observations $Z(\mathbf{x}_i)$}
%\label{fig:local neighborhood}
%\end{figure}

A \emph{global} method makes use of all available observations in the region of interest (e.g. the image of a whole portion of the sky) to derive the estimated value at the target point whereas a \emph{local} method only considers observations found within some small neighborhood around the target point. Thus, global methods may be preferable to capture the general trend in the data, whereas local methods may better capture the local or short-range variations and exploit spatial relationships in the data \citep{Burrough1998}. A trend surface is an example of global estimator.\newline\indent An interpolation methods that produces an estimate that is the same as the observed value at a sampled point is called an \emph{exact} method. On the contrary a method is \emph{approximate} if its predicted value at the point differs from its known value: some amount of smoothing is involved for avoiding sharp peaks or troughs in the resulting fitted surface. \newline\indent Lastly, a \emph{stochastic} (also called \emph{geostatistical}) interpolator incorporates the concept of randomness and yields both an estimated value (the deterministic part) and an associated error (the stochastic part, e.g. an estimated variance). On the other hand, a \emph{deterministic} method does not provide any assessment of the error made on the interpolated value. \newline\indent Table~\ref{table:interpolation methods} contains the list of spatial interpolation methods  covered in this article along with their classification.\newline

\begin{figure}
\scalebox{0.57}{\includegraphics[trim=10mm 10mm 10mm 10mm, clip]{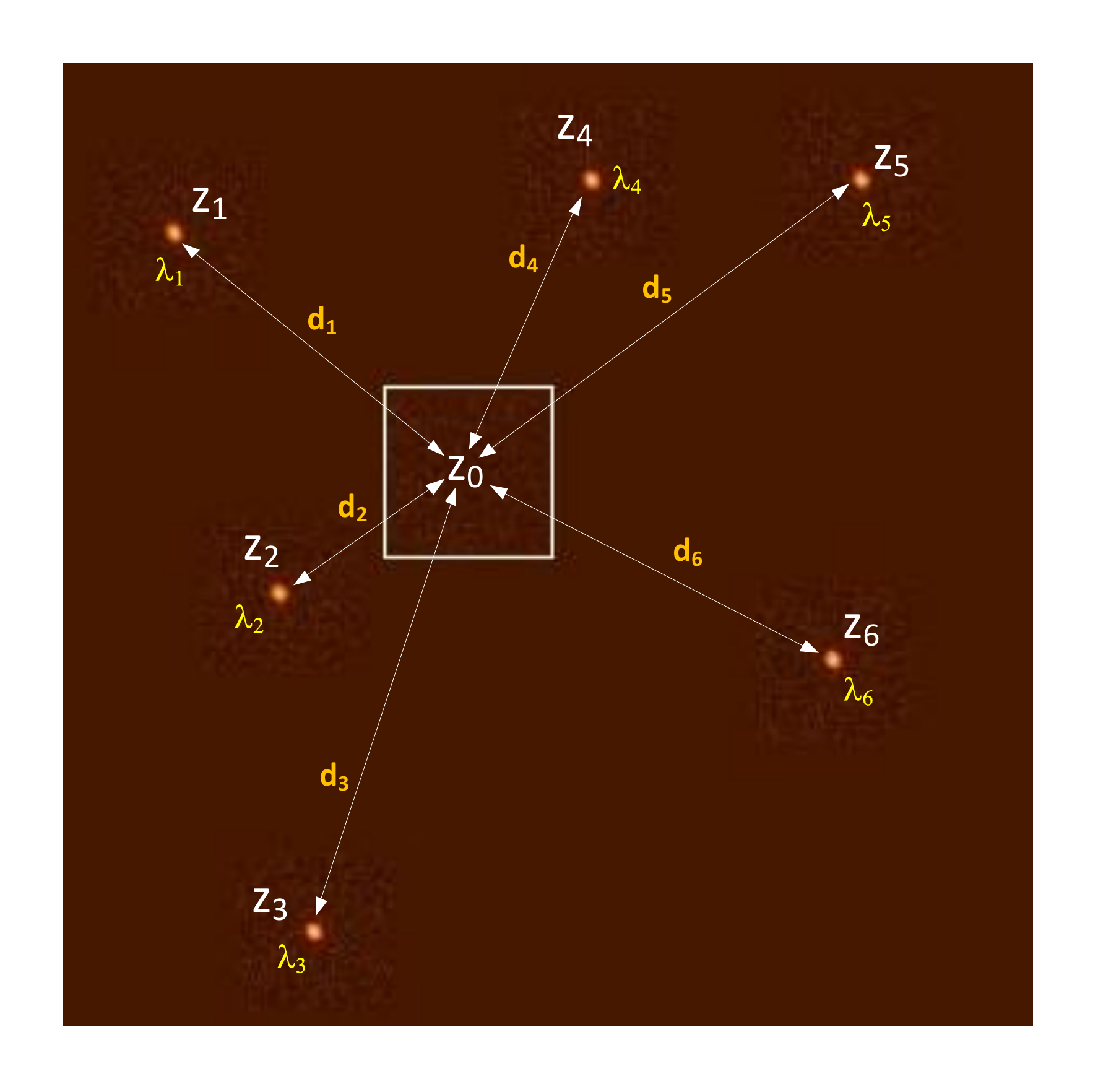}} 
%\resizebox{\hsize}{!}{\scalebox{0.30}{\includegraphics[trim=10mm 10mm 10mm 10mm, clip]{fig_local_weighted_interpolation.pdf}} }
%\resizebox{\hsize}{!}{\includegraphics{fig_local_weighted_interpolation.pdf}}
\caption{An illustration of local interpolation between a set of \mbox{neighboring observations} $Z(\mathbf{x}_i)$ at distances $d_i$ from a target location $\mathbf{x}_0$. In this example, a set of weights $\lambda_i$ is assigned to each of the $Z(\mathbf{x}_i)$, as in Eq.~\ref{eq:generic interpolation formula}}
\label{fig:local weighted interpolation}
\end{figure}

Nearly all methods of spatial interpolation share the following general spatial prediction formula
\begin{equation}\label{eq:generic interpolation formula} \hat z(\mathbf{x}_0)=\sum_{i=1}^N{\lambda_i\, z(\mathbf{x}_i)}\end{equation}
where $\mathbf{x}_0$ is a target point where the value should be estimated, the $z(\mathbf{x}_i)$ are the locations where an observation is available and the $\lambda_i$ are the weights assigned to individual observations. $N$ represents the number of points involved in the estimation (See Fig.~\ref{fig:local weighted interpolation} for an illustration). Each interpolation method has its own algorithm for estimating the weights $\lambda_i$. All the interpolation methods evaluated in this article except splines, follow Eq.~(\ref{eq:generic interpolation formula}).\\	

We now review several widely used interpolation schemes that can be applied to the PSF interpolation problem: polynomial splines, inverse distance weighting (IDW), radial basis functions (RBF) and Kriging. In the remaining sections, we test these interpolation methods using the GREAT10 Star Challenge simulated data.

\begin{table}
\renewcommand{\arraystretch}{1.25}
\caption{Spatial interpolation methods reviewed in this article}              % title of Table
\label{table:interpolation methods}      % is used to refer this table in the text
\centering                                      % used for centering tableto avoid sharp peaks or troughs in the output surface
\begin{tabular}{p{3.5cm} l p{1.5cm} l}          % centered columns (4 columns)
\hline                                             %inserts single line
\hline                       % inserts double horizontal lines
Interpolation method & Scope & Exactness & Model \\    % table heading
\hline                                   % inserts single horizontal line
   Polynomial fitting & global & approximate & deterministic \\      % inserting body of the table
   Basis splines & global & approximate\tablefootmark{1} & deterministic \\
   Inverse distance weighting & local & exact\tablefootmark{2} & deterministic \\
   Radial basis function & local & exact\tablefootmark{3} & deterministic \\
   Ordinary Kriging & local & exact\tablefootmark{4} & stochastic \\
\hline                                             %inserts single line
\end{tabular}
\tablefoot{
\tablefoottext{1}{Can be made exact by disabling smoothing} \tablefoottext{2}{Smoothing possible with specific algorithms} \tablefoottext{3}{Some Kriging algorithms are approximate}
}
\end{table}

\subsection{Spline interpolation}
\label{subsection:spline interpolation}
A (polynomial) univariate \emph{spline} or degree $p$ (order $p+1$) is made of a set of polynomial pieces, joined together such that pieces and their derivatives at junction points (\emph{knots}) are continuous up to degree $p-1$ \citep{Deboor1978, Dierckx1995, Schumaker2007, Prenter2008}.\newline\indent When it comes to modeling two-dimensional spatially varying PSF attributes across an image, we are more specifically interested in bivariate polynomial splines. A function $s(x,y)$ defined on a domain $[a,b]\,\times\,[c,d]$ with respective, strictly increasing knot sequences $\lambda_i$, $i=0,1...,g+1$ $(\lambda_0=a, \lambda_{g+1}=b)$ in the $x$ direction and $\mu_j$, $j=0,1...,h+1$  $(\mu_0=c, \mu_{h+1}=d)$ in the $y$ direction is called a bivariate (tensor product) spline function of degree $k>0$ (order $k+1$) in $x$ and $l>0$ (order $l+1$) in $y$ if the following two conditions are satisfied: 
\begin{enumerate}
\item On each subregion $\mathcal{D}_{i,j}=[\lambda_i,\lambda_{i+1}]\,\times\,[\mu_j,\mu_{j+1}]$, $s(x,y)$ is given by a polynomial of degree $k$ in $x$ and $l$ in $y$ \begin{displaymath}\label{eq:spline def 1}s(x,y)\in  \mathcal{P}_k \otimes \mathcal{P}_l \quad i=0,1,...,g; j=0,1,...h\end{displaymath}
\item The function $s(x,y)$ and all its partial derivatives are continuous on $\mathcal{D}_{i,j}$\begin{displaymath}\label{eq:spline def 2}\frac{\partial^{i+j}s(x,y)}{\partial x^i\,\partial y^j}\in \mathcal{C}(\mathcal{D}_{i,j}) \quad i=0,1,...,k-1; j=0,1,...,l-1\end{displaymath}
\end{enumerate}
We saw earlier that polynomial fitting suffers in particular from two serious drawbacks. One of these is that individual observations can exert an influence, in unexpected ways, on different parts of the interpolating surface. The other is the tendency of the interpolation surface to wiggle without control as soon as one increases the degree of the polynomial to try to obtain a closer fit. 

Polynomial splines solve these problems in two ways. First, a spline is not made of a single "global" polynomial but of a set of "local" polynomial pieces. This design confines the influence of individual observations within the area covered by the enclosing polynomial piece. In most applications, a specific type of spline is preferred, the so-called "Basis spline" (\emph{B-spline}), built from as a linear combination of basis polynomial functions called \emph{B-splines}\begin{displaymath}\label{eq:b-spline def}s(x,y)=\sum_{i=-k}^g\sum_{j=-l}^hc_{i,j}\,N_{i,k+1}(x)\,M_{j,l+1}(y)\end{displaymath} where $N_{i,k+1}(x)$ and $M_{j,l+1}(y)$ are B-splines defined on the $\lambda$ and $\mu$ knot sequences respectively. B-splines are popular for their computational efficiency, e.g. with the algorithms of Cox \citep{Cox1972} or de Boor \citep{DeBoor1972}. For a formal definition of the B-spline basis see e.g. \citet{Deboor1978, Dierckx1995, Prenter2008}.

The second issue is solved by the ability to control the smoothness of the spline. The example of polynomial fitting shows that a good fit to the data is not the one and only goal in surface fitting; another, and conflicting, goal is to obtain an estimate that does not display spurious fluctuations.  A successful interpolation is, actually, a tradeoff between goodness of fit (fidelity to the data) and roughness (or "wiggleness") of fit: a good balance between these two criteria will allow the approximation to not pick up too much noise (overfitting) while avoiding signal loss (underfitting).

There is an extensive literature on spline interpolation and many algorithms and variants have been developed since their invention in the 1960s. Still, one can divide spline interpolation algorithms into two main families: those based on the so-called \emph{constructive} approach, where the form of the spline function is specified in advance and the estimation problem is reduced to the determination of a discrete set of parameters; and those that follow a \emph{variational} approach, where the approximation function is not known a priori, but follows from the solution of a variational problem, which can often be interpreted as the minimization of potential energy. We outline both approaches below.

\subsubsection*{Variational approach of spline interpolation}
The variational approach \citep{Wahba1990, Green1994} consists in minimizing the functional\begin{equation}\label{eq:spline variational functional}S(f,\alpha)=\sum_{i=1}^N{\Arrowvert z(\mathbf{s}_i)-f(\mathbf{s}_i)}\Arrowvert^2+\alpha\int_\mathcal{D}\{{f^{(m)}}\}^2d{\mathbf{s}_i}\end{equation} where the bivariate spline function $f$ is fitted to the $z(\mathbf{s}_i), i=0,...,N$ set of points in the region $\mathcal{D}$ where the approximation is to be made. It can be shown, e.g. \citep{Green1994} that the solution is a natural spline, that is, a spline whose second and third derivatives are zero at the boundaries. splines obtained in such a way as known in the literature as \emph{smoothing splines}. The parameter $m$ represents  the order of the derivative of $f$ and $\alpha\geq0$ is a smoothing parameter controlling the tradeoff between fidelity to the data and roughness of the spline approximation.\begin{enumerate}
\item As $\alpha \longrightarrow 0$ (no smoothing), the left-hand side least squares estimate term dominates the roughness term on the right-hand side and the spline function attempts to match every single observation point (oscillating as required)
\item As $\alpha \longrightarrow \infty$ (infinite smoothing), the roughness penalty term on the right-hand side becomes paramount and the estimate converges to a least squares estimate at the risk of underfitting the data.
\end{enumerate}
The most popular variational spline interpolation scheme is that based on the \emph{thin-plate spline} (TPS) \citep{Duchon1976, Meinguet1979, WahbaWendelberger1980, Wahba1990, Hutchinson1995}. The TPS interpolating spline is obtained by minimizing an energy function of the form (\ref{eq:spline variational functional})  \begin{equation}\label{eq:TPS spline}S(f,\alpha)=\sum_{i=1}^N{\Arrowvert z(\mathbf{s}_i)-f(\mathbf{s}_i)}\Arrowvert^2+\alpha\,J_m(g)\,d{\mathbf{s}_i}\end{equation}
The most common choice of $m$ is 2 with $J_2$ of the form\begin{equation}\label{eq:TPS spline partials}J_2(g)=\int_\mathcal{D}\big\{\big(\frac{\partial ^2g}{\partial x^2}\big)^2+2\big(\frac{\partial ^2g}{\partial x\,\partial y}\big)^2+\big(\frac{\partial ^2g}{\partial y^2}\big)^2\big\}\,dx\,dy\end{equation} where the roughness function $g(x,y)$ is given by\begin{equation}g(\mathbf{s})=a_0+a_1x+a_2y+\sum_1^N\lambda_i\,\phi(\mathbf{s}-\mathbf{s}_i)\end{equation} $\phi$ being the radial basis function (RBF): $\Phi(x,y)=d_i^2\,ln(d_i)$ with euclidean distance $d_i=\sqrt{(x-x_i)^2+(y-y_i)^2}$. The $\lambda_i$ are weighting factors.

\subsubsection*{Constructive approach of spline interpolation}

Interpolating splines obtained through such a scheme are often referred to as \emph{least squares splines} \citep{Dierckx1980, HayesHalliday1994,Dierckx1995}. For such splines, goodness of fit is measured through a least squares criterion, as in the variational approach, but smoothing is implemented in a different way: in the variational solution, the number and positions of knots are not varied, and the approximating spline is obtained by minimizing an energy function. On the other hand, in the constructive approach, one still tries to find the smoothest spline that is also closest to the observation points. But this is achieved by optimizing the number and placement of the knots and finding the corresponding coefficients $\mathbf{c}$ in the B-Spline basis. This is measured by a so-called \emph{smoothing norm} $G(\mathbf{c})$. Thus, the approximating spline arises as the minimization of 
\begin{equation}
\label{eq:constructive spline}
S(f,\alpha)=\sum_{i=1}^N{\Arrowvert z(\mathbf{s}_i)-f(\mathbf{s}_i)}\Arrowvert^2+\alpha\,G(\mathbf{c})
\end{equation} 
using the same notation as in (\ref{eq:spline variational functional}). An example of knot placement strategy is to increase the number of knots (i.e. reduce the inter-knot distance) in areas where the surface to fit varies faster or more abruptly. By the way, we note that minimization is not obtained by increasing the degree of the spline (which is kept low, typically 3).

Whatever the approach followed for obtaining a suitable interpolating spline, spline interpolation is essentially \emph{global}, \emph{approximate} and \emph{deterministic}, as it involves all available observations points, makes use of smoothing and does not provide any estimation on interpolation errors. The interpolation can however be made exact by setting the smoothing coefficient to zero. Also, for smoothing splines (variational approach) a technique called \emph{generalized cross-validation} (GCV) \citep{CravenWahba1979, Wahba1990, HutchinsonGessler1994} allows to automatically choose, in expression (\ref{eq:TPS spline}), suitable parameters for $\alpha$ and $m$ for minimizing cross-validation residuals. Otherwise, one can always use standard cross-validation or Jackknifing to optimize the choice of input parameters (see Sect.~\ref{subsection:cross-validation}).

The most frequently used splines for interpolation are \mbox{cubic} splines, which are made of polynomials pieces of degree at most 3 that are twice continuously differentiable. Experience has shown that in most applications, using splines of higher degree seldom yields any advantage. As we saw earlier, splines avoid the pitfalls of polynomial fitting while being much more flexible, which allows them, despite their low degree, to capture finer-grained details. The method assumes the existence of measurement errors in the data and those can be handled by adjusting the amount of smoothing.

On the minus side, cubic or higher degree splines are sometimes criticized for producing an interpolation that is "too smooth". They also keep a tendency to oscillate (although this can be controlled unlike with standard polynomials). In addition, the final spline estimate is influenced by the number and placement of knots, which confers some arbitrariness to the method, depending on the approach and algorithm used. This can be a problem since there is, in general, no built-in mechanism for quantifying interpolation errors. Lastly, spline interpolation is a global method and performance may suffer on large datasets. A summary of the main strengths and weaknesses of spline interpolation is given in Table~\ref{table:spline interpolation}.

\begin{table}
\renewcommand{\arraystretch}{1.25}
\caption{Spline interpolation: Pros and cons}              % title of Table
\label{table:spline interpolation}      % is used to refer this table in the text
%\centering                                      % used for centering table
\begin{tabular}{l p{7.5cm}}          % centered columns (4 columns)
\hline\hline                        % inserts double horizontal lines
\multicolumn{2}{l}{Spline Interpolation} \\    % table heading
\hline                                   % inserts single horizontal line
\multirow{2}{*}{Pros} & Able to capture both broad and detailed features  \\ 
                                      & The tradeoff between goodness and roughness of fit can be adjusted through smoothing \\ 
\hline 
\multirow{4}{*}{Cons}  & Overall smoothness may still be too high \\
                       & Keep a tendency to oscillate \\ 
                       & No estimation of interpolation errors in most algorithms \\ 
                       & Potentially less efficient to compute than local interpolation algorithms \\ 
\hline                                             %inserts single line
\end{tabular}
\end{table}

\subsection{inverse distance weighting}
\label{subsection:IDW}
Inverse distance weighting (IDW) \citep{Shepard1968} is one of the oldest spatial interpolation method but also one of the most commonly used. The estimated value $\hat z(\mathbf{x}_0)$ at a target point $\mathbf{x}$ is given by Eq.~(\ref{eq:generic interpolation formula}) where the weights $\lambda_i$ are of the form: 
\begin{equation}\label{eq:IDW weights}\lambda_i=\frac{1}{d^{\, \beta}(\mathbf{x}_0,\mathbf{x}_i)}\, \big/  \sum_{i=1}^N {\frac{1}{d^{\, \beta}(\mathbf{x}_0,\mathbf{x}_i)}} \qquad \beta \geq 0 \qquad \sum_{i=1}^N \lambda_i=1 \end{equation}
In the above expression, $d(\mathbf{x}_0,\mathbf{x}_i)$ is the distance between points $\mathbf{x}_0$ and $\mathbf{x}_i$, $\beta$ is a power parameter and $N$ is the number of points found in some neighborhood around the target point $\mathbf{x}_0$. Scaling the weights $\lambda_i$ so that they sum to unity ensures the estimation is unbiased.

The rationale behind this formula is that data points near the target points carry a larger weight than those further away. The weighting power $\beta$ determines how fast the weights tend to zero as the distance $d(\mathbf{x}_0,\mathbf{x}_i)$ increases. That is, as $\beta$ is increased, the predictions become more similar to the closest observations and peaks in the interpolation surface becomes sharper. In this sense, the $\beta$ parameter controls the degree of smoothing desired in the interpolation.

Power parameters between $1$ and $4$ are typically chosen and the most popular choice is $\beta=2$, which gives the inverse-distance-squared interpolator. IDW is referred to as "moving average" when $\beta=0$ and "linear interpolation" when $\beta=1$.

For a more detailed discussion on the effect of the power parameter $\beta$, see e.g. \citet{Laslett1987, Burrough1988, Brus1996, CollinsBolstad1996}. Another way to control the smoothness of the interpolation is to vary the size of the neighborhood: increasing $N$ yields greater smoothing.\newline

IDW is a \emph{local} interpolation technique because the estimation at $\mathbf{x}_0$ is based solely on observations points located in the neighboring region around $\mathbf{x}_0$ and because the influence of points further away decreases rapidly for $\beta > 0$. It is also forced to be \emph{exact} by design since the expression for $\lambda_i$ in Eq.~(\ref{eq:IDW weights}) reaches the indeterminate form $\infty/\infty$ when the estimation takes place at the point $\mathbf{x}_0$ itself. IDW is further labeled as \emph{deterministic} because the estimation algorithm relies purely on geometry (distances) and does not provide any estimate on the error made.

%IDW is a \emph{local} interpolation technique because the estimation at $\mathbf{x}_0$ is based solely on observations points located in the neighboring region around $\mathbf{x}_0$ and because the influence of points further away decreases rapidly for $\beta > 0$. It is also forced to be \emph{exact} by design since $\sum_{i=1}^N 1/{{d(\mathbf{x}_0,\mathbf{x}_i)}\,^\beta}$ in Eq.~(\ref{eq:IDW weights}) tends to infinity when the estimation takes place at the point $\mathbf{x}_0$ itself. IDW is further labeled as \emph{deterministic} because the estimation algorithm relies purely on geometry (distances) and does not provide any estimate on the error made.

IDW is popular for its simplicity, computational speed and ability to work on scattered data. The method also has a number of drawbacks. One is that the choice of the $\beta$ parameter and the \mbox{neighborhood} size and shape are arbitrary, although techniques such as cross-validation or jackknifing can provide hints for tuning these parameters (see \ref{subsection:cross-validation}). Another is that there exists no underlying statistical model for measuring uncertainty in the predictions. Further, the results of the method method are sensitive to outliers and influenced by the way observations have been sampled. In particular, the presence of clustering can bias the estimation since in such cases clustered points and isolated points at similar but opposite distances will carry about the same weights. A common feature of IDW-generated interpolation surfaces is the presence of spikes or pits around observation points since isolated points have a marked influence on the prediction in their vicinity.

The original Shepard algorithm has been enhanced by several authors to address some of the shortcomings listed above. See in particular \citet{Renka1988}, \citet{Tomczak1998} and \citet{Lukaszyk2004}. One frequent extension consists in explicitly introducing a smoothing factor $s$ into Eq.~(\ref{eq:IDW weights}), which then becomes

\begin{equation}
\label{eq:IDW weights smoothing}
\lambda_i=\frac{1}{{\big(d(\mathbf{x}_0,\mathbf{x}_i) + s}\big)^{\, \beta}}\, \big/  \sum_{i=1}^N {\frac{1}{{\big(d(\mathbf{x}_0,\mathbf{x}_i) + s}\big)^{\,\beta} }  }
\end{equation} 
with values of $s$ typically chosen between $1$ and $5$. Table~\ref{table:IDW} summarizes the main pros and cons of inverse distance weighting.

\begin{table}
\renewcommand{\arraystretch}{1.25}
\caption{inverse distance weighting: Pros and cons}              % title of Table
\label{table:IDW}      % is used to refer this table in the text
\begin{tabular}{l p{7.5cm}}          % centered columns (4 columns)
\hline\hline                        % inserts double horizontal lines
\multicolumn{2}{l}{inverse distance weighting} \\    % table heading
\hline                                   % inserts single horizontal line
\multirow{2}{*}{Pros} & Simple and intuitive \\ 
                      & Fast to compute \\ 
\hline 
\multirow{5}{*}{Cons} & Choice of interpolation parameters empirical \\
                      & The interpolation is always exact (no smoothing) \\
                      & Sensitivity to outliers and sampling configuration (clustered and isolated points) \\
                      & No estimation of interpolation errors (deterministic) \\ 
\hline                                             %inserts single line
\end{tabular}
\end{table}

\subsection{Interpolation with radial basis functions}
\label{subsection RBF interpolation}
We just described IDW, a simple form of interpolation on scattered data where the weighting power ascribed to a set neighboring point $x_i$ from some point $x$ only depends on an inverse squared distance function.\newline\indent
We now describe a similar, but more versatile form of interpolation where the distance function is more general and expressed in terms of a \emph{radial basis function} (RBF) \citep{Buhmann2003, NumericalRecipes2007}. A RBF function, or \emph{kernel} $\phi$ is a real-valued function where the value evaluated at some point $\mathbf{x}_0$ only depends on the radial distance between $\mathbf{x}_0$ and a set of points $\mathbf{x}_i$, so that $\phi(\mathbf{x}_0-\mathbf{x}_i)=\phi(\Arrowvert \mathbf{x}_0-\mathbf{x}_i\Arrowvert)$. The norm usually represents the Euclidean distance but other types of distance functions are also possible.\newline\indent 
The idea behind RBF interpolation is to consider that the influence of each observation on its surrounding is the same in all direction and well described by a RBF kernel. The interpolated value at a point $\mathbf{x}_0$ is a weighted linear combination of RBF evaluated on points located within a given neighborhood of size $N$ according to the expression \begin{equation}\label{eq:RBFinterpolation}\hat z(\mathbf{x}_0)=\sum_{i=1}^N{\lambda_i\, z(\mathbf{x}_i)}=\sum_{i=1}^N{\lambda_i\,\phi(\Arrowvert (\mathbf{x}_0-\mathbf{x}_i}\Arrowvert)\end{equation} The weights are determined by imposing that the interpolation be exact at all neighboring points $\mathbf{x}_i$, which entails the resolution of a linear system of $N$ equations with $N$ unknown weighting factors $\lambda_i$.  In some cases, it is necessary to add a low-degree polynomial $P_k(\mathbf{x})$ of degree $k$ to account for a trend in $z(\mathbf{x})$ and ensure positive-definiteness of the solution. Expression (\ref{eq:RBFinterpolation}) is then transformed into  \begin{equation}\label{eq:RBFinterpolation poly}\hat z(\mathbf{x}_0)=P_k(\mathbf{x}_0)+\sum_{i=1}^N{\lambda_i\,\phi(\Arrowvert (\mathbf{x}_0-\mathbf{x}_i}\Arrowvert)\end{equation} Sometimes, an interpolation scheme based on a \emph{normalized RBF} (NRBF) of the form  \begin{equation}\label{eq:RBFinterpolation normalized}\hat z(\mathbf{x}_0)=\sum_{i=1}^N{\lambda_i\,\phi(\Arrowvert (\mathbf{x}_0-\mathbf{x}_i}\Arrowvert)\, \big/ \sum_{i=1}^N{\phi(\Arrowvert (\mathbf{x}_0-\mathbf{x}_i}\Arrowvert)\end{equation} is preferred to (\ref{eq:RBFinterpolation}), although no significant evidence for superior performance has been found.\newline

The actual behavior and accuracy of RBF interpolation closely depends on how well the $\phi$ kernel matches the spatial distribution of the data. The most frequently used RBF kernels are listed in Table~\ref{table:popular RBF kernels}, where $r=\Arrowvert\mathbf{x}-\mathbf{x}_i\Arrowvert$ and the quantity $\epsilon$ is the so-called \emph{shape parameter}. The required conditions for $\phi$ to be a suitable RBF kernel have been given by \citet{Micchelli1986} but the choice of the most adequate kernel for a problem at hand is often empirical.\newline\indent
The shape parameter $\epsilon$ contained in the Multiquadric, Inverse Multiquadric and Gaussian kernels influences the shape of the kernel function and controls the tradeoff between fitting accuracy and numerical stability. A small shape parameter produces the most accurate results, but is always associated with a poorly conditioned interpolation matrix. Despite the research work of e.g. \citet{Hardy1990}, \citet{Foley1994} and \citet{Rippa1999}, finding the most suitable shape parameter is often a matter of trial and error. A rule of thumb is to set $\epsilon$ to approximately the mean distance to the nearest neighbor. \newline\indent
RBF interpolation based on the \emph{Multiquadric} (MQ) kernel $\sqrt{1+(\epsilon\,r)^2}$ is the most common. It was first introduced by \citet{Hardy1971} as a "superpositioning of quadric surfaces" for solving a problem in cartography. In its review of interpolation methods on scattered data, \citet{Franke1982} highlighted the good performance of the MQ kernel, which has, since then proven highly successful in many disciplines \citep{Hardy1990}.

\begin{table}[t]
\renewcommand{\arraystretch}{1.25}
\caption{Most popular RBF kernels}              % title of Table
\label{table:popular RBF kernels}      % is used to refer this table in the text
%\centering                                      % used for centering table
\begin{tabular}{l l}          % centered columns (4 columns)
\hline\hline                        % inserts double horizontal lines
RBF kernel $\phi(r)$ & Expression\\    % table heading
\hline                                   % inserts single horizontal line
    Multiquadric &  $\sqrt{1+(\epsilon\,r)^2}$  \\[0.5mm]      % inserting body of the table
    Inverse Multiquadric & $1/[1+(\epsilon\,r)^2]$ \\[0.5mm]
    Gaussian & $exp[-(\epsilon\,r)^2]$ \\[0.5mm]
    Thin-Plate & $r^2\,ln(r)$ \\[0.5mm]
    Linear & $r$\\[0.5mm]
    Cubic & $r^3$ \\[0.5mm]
\hline                                             %inserts single line
\end{tabular}
\end{table}

RBF interpolation is fundamentally a \emph{local}, \emph{exact} and \emph{deterministic} method. There are, however, algorithms that allow to introduce smoothing to better handle noise and measurement errors in the data. The method can prove highly accurate but this really depends on the affinity between the data and the kernel function used. Also, because predictions are exact, RBF functions can be locally sensitive to outliers. As for other deterministic methods like splines or IDW, the optimal set of parameters are most often determined by cross-validation or Jackknifing (see Sect.~\ref{subsection:cross-validation}). Table~\ref{table:RBF} recapitulates the favorable and less favorable aspects of interpolation based on radial basis functions.

\begin{table}[t]
\renewcommand{\arraystretch}{1.25}
\caption{radial basis functions for interpolation: Pros and cons}              % title of Table
\label{table:RBF}      % is used to refer this table in the text
\begin{tabular}{l p{7.5cm}}          % centered columns (4 columns)
\hline\hline                        % inserts double horizontal lines
\multicolumn{2}{l}{radial basis functions} \\    % table heading
\hline                                   % inserts single horizontal line
\multirow{2}{*}{Pros} & Flexibility, thanks to various choice of kernel functions \\ 
                      & Relatively fast (local method), but computational speed depends on the kernel function \\ 
\hline 
\multirow{4}{*}{Cons} & Choice of kernel functions and interpolation parameters \mbox{empirical} \\
                      & The interpolation is always exact (no smoothing) \\
                      & Sensitivity to outliers and sampling configuration (clustered and isolated points) \\
                      & No estimation of interpolation errors (deterministic) \\ 
\hline                                             %inserts single line
\end{tabular}
\end{table}

\subsection{Kriging}
\label{subsection:kriging}

Kriging is a spatial prediction technique initially created in the early 1950's by mining engineer Daniel G. Krige \citep{Krige1951} with the intent of improving ore reserve estimation in South Africa. But it was essentially the mathematician and geologist Georges Matheron who put Krige's work a firm theoretical basis and developed most of the modern Kriging formalism \citep{Matheron1962, Matheron1963}. 

Following Matheron's work, the method has spread from mining to disciplines such as hydrology, meteorology or medicine, which triggered the creation of several Kriging  \mbox{variants}. It is thus more accurate to refer to Kriging as a family of spatial prediction techniques instead of a single method. It is also essential to understand that Kriging constitutes a general method of interpolation that is in principle  applicable to any discipline, such as astronomy.

The following textbooks provide a good introduction to the subject: \citet{Journel1978, IsaaksSrivastava1989, Cressie1991, ClaytonDeutschJournel1997, Goovaerts1997, Chiles1999, Wackernagel2003, Waller2004, WebsterOliver2007}.

Like most of the local interpolation methods described so far in this article, Kriging makes use of the weighted sum ($\ref{eq:generic interpolation formula}$) to estimate the value at a given location based on nearby observations. But instead of computing weights based on geometrical distances only, Kriging also takes into account the  \mbox{\emph{spatial correlation}} existing in the data. It does so by treating observed values $z(\mathbf{x})$ as \emph{random variables} $Z(\mathbf{x})$ varying according to a spatial \emph{random process}~\footnote{also called \emph{random function} or \emph{stochastic process}}. In fact, Kriging assumes the underlying process has a form of second-order stationarity called \mbox{\emph{intrinsic stationarity}}. Second-order stationarity is traditionally defined as follows:
%A second-order stationary random process is charaterized by a constant mean and a covariance that only depends on the separation distances and directions between two locations. In other words,
\begin{enumerate}
\item The mathematical expectation $E(Z(\mathbf{x}))$ exists and  does not depend on $\mathbf{x}$ \begin{equation}\label{eq:expectation 2nd stationarity}E\big[Z(\mathbf{x})\big]=m,\quad\forall{\,\mathbf{x}}\end{equation}
\item For each pair of random variable $\big\{Z(\mathbf{x}), Z(\mathbf{x+h}) \big\}$, the covariance exists and only depends on the separation vector $\mathbf{h}=\mathbf{x}_j-\mathbf{x}_i$, \begin{equation}\label{eq:covariance 2nd stationarity}C(\mathbf{h})=E\big\{\big[Z(\mathbf{x+h})-m\big] \, \big[Z(\mathbf{x})-m\big] \big\}, \quad\forall{\,\mathbf{x}}\end{equation}
\end{enumerate}
Kriging's \emph{intrinsic stationarity} \citep{Matheron1963, Matheron1965} is a slightly weaker form of second-order stationarity where the difference $Z(\mathbf{x+h})-Z(\mathbf{x})$ is treated as the stationary variable instead of $Z(\mathbf{x})$:
\begin{enumerate}
\item \begin{equation}\label{eq:expectation intrinsic stationarity}E\big[Z(\mathbf{x+h})-Z(\mathbf{x})\big]=0,\quad\forall{\,\mathbf{x}}\end{equation}
\item \begin{equation}\label{eq:semivariance}Var\big[Z(\mathbf{x+h})-Z(\mathbf{x})\big]=E\big\{{\big[Z(\mathbf{x+h})-Z(\mathbf{x})\big]}^2 \big\}=2\gamma(\mathbf{h})\end{equation}
\end{enumerate} The function $\gamma(\mathbf{h})$ is called \emph{semivariance} and its graph \emph{semivariogram} or simply \emph{variogram}.\newline\indent
One reason for preferring intrinsic stationarity over secondary stationarity is that semivariance remains valid under a wider range of circumstances. When covariance exists, both stationarities are related through \begin{equation}\label{eq:equivalence covariance-variogram}\gamma(\mathbf{h})=C(0)-C(\mathbf{h}), \qquad C(0)=Var\big[Z(\mathbf{x})\big]\end{equation}

\begin{figure}
%\scalebox{0.75}{\includegraphics{fig_variogram_covariance.pdf}} 
\resizebox{\hsize}{!}{\includegraphics[trim=5mm 5mm 5mm 5mm, clip]{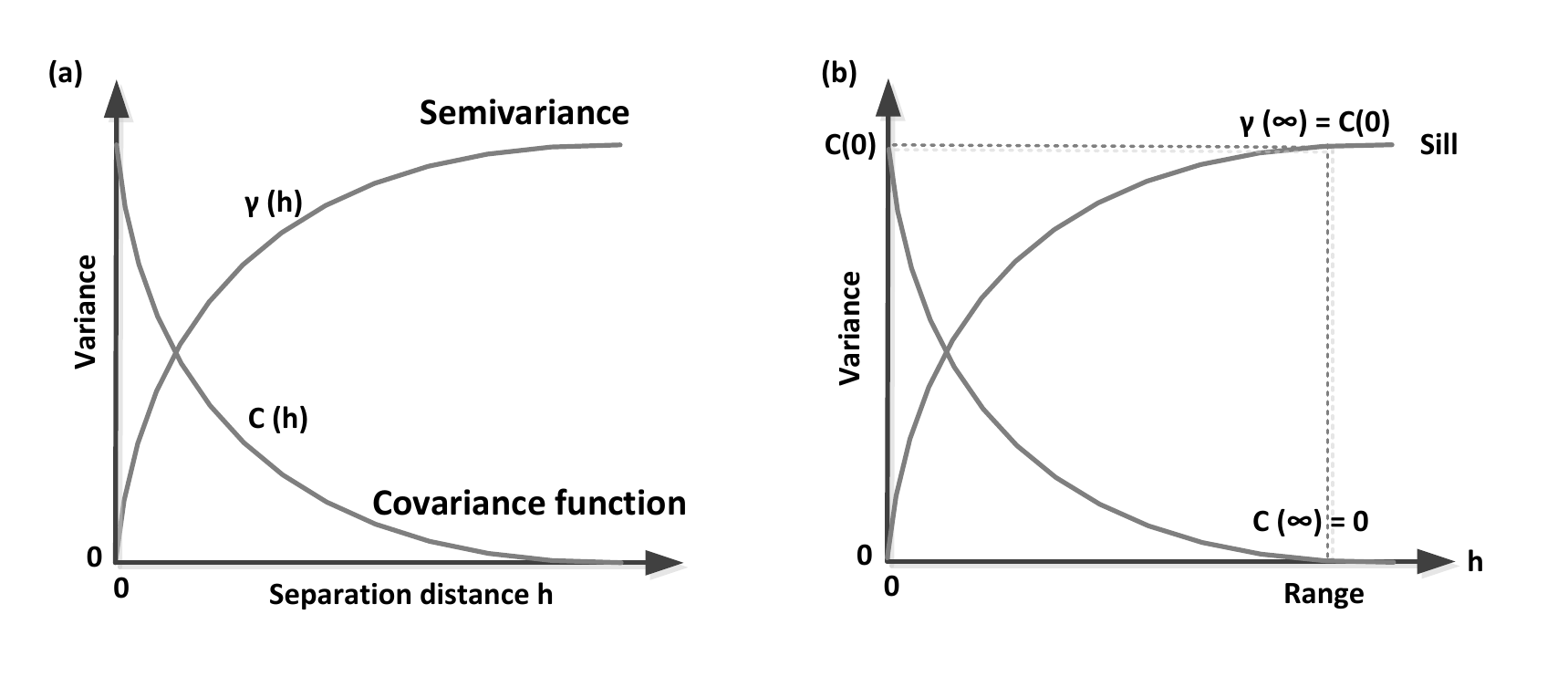}}
\caption{\textbf{(a)} typical variogram $\gamma(\mathbf{h})$ and its equivalent covariance function $C(\mathbf{h})$: if the data has some sort of spatial autocorrelation, nearby (small $h$) $Z(\mathbf{x}$) observed values will be more similar than more distant $Z(\mathbf{x})$ values (larger $h$); \textbf{(b)} as the separation distance $h$ grows, the quantity $Z(\mathbf{x+h})-Z(\mathbf{x})$ in expression (\ref{eq:semivariance}) will tend to increase on average, but less and less as the influence of $Z(\mathbf{h})$ on $Z(\mathbf{x+h})$ weakens; at some threshold distance $h$, called the \emph{range}, the increase in variance becomes negligible and the asymptotical variance value is known as the \emph{sill}}
\label{fig:variogram covariance function with sill and lag}
\end{figure}

Fig.~\ref{fig:variogram covariance function with sill and lag} shows a typical variogram along with its equivalent covariance function.

%Fig.~\ref{fig:variogram covariance function with sill and lag}~(a) shows a typical variogram along with its equivalent covariance function. If the data has some sort of spatial autocorrelation, nearby (small $h$) $Z(\mathbf{x}$) observed values will be more similar than more distant $Z(\mathbf{x})$ values (larger $h$). That is, the variance between close values will be lower (higher covariance) than that between remote values (lower covariance). This type of variogram is referred to as a \mbox{\emph{bounded variogram}} and is illustrated in Fig.~\ref{fig:variogram covariance function with sill and lag}~(b).\newline

Over the years about a dozen Kriging variants have been developed. We will concentrate here on \emph{ordinary Kriging} (OK), which is, by far, the most widely used. The description of other forms of Kriging can be found in the literature given at the beginning of this section.

ordinary Kriging is a \emph{local}, \emph{exact} and \emph{stochastic} method. The set of $Z(\mathbf{x})$ is assumed to be an intrinsically stationary random process of the form
\begin{equation}\label{eq:OK random process}Z(\mathbf{x})=m+\epsilon\,({\mathbf{x}})\end{equation} 
The quantity $\epsilon({\mathbf{x}})$ is a random component drawn from a probability distribution with mean zero and variogram $\gamma(\mathbf{h})$ given by~(\ref{eq:semivariance}). The mean $m=E[Z(\mathbf{x})]$ is assumed constant because of (\ref{eq:expectation intrinsic stationarity}), but remains \emph{unknown}. The ordinary Kriging predictor is given by the weighted sum 
\begin{equation}\label{eq:generic interpolation formula 2}\hat Z(\mathbf{x}_0)=\sum_{i=1}^N{\lambda_i\, Z(\mathbf{x}_i)}\end{equation}
where the weights $\lambda_i$ are obtained by minimizing the so-called \emph{Kriging variance}
\begin{equation}\label{eq:kriging variance}{\sigma^2(\mathbf{x}_0)}=Var\big[\hat Z(\mathbf{x}_0)-Z(\mathbf{x}_0)\big]=E\big\{{\big[\hat Z(\mathbf{x}_0)-Z(\mathbf{x}_0)\big]}^2 \big\}\end{equation} 
subject to the unbiaseness condition 
\begin{equation}\label{eq:kriging unbiaseness condition}{E\big[\hat Z(\mathbf{x}_0)-Z(\mathbf{x}_0)\big]}=0={\sum_{i=1}^N{\lambda_i\, E\big[z(\mathbf{x}_i)\big]}-m}\end{equation}
%Minimizing (\ref{eq:kriging variance}) yields $\sum_{i=1}^N{\lambda_i\, \gamma(\mathbf{x}_i,\mathbf{x}_j)+\mu=\gamma(\mathbf{x}_i,\mathbf{x}_0)}$ where $\mu$ is a Lagrange multiplier, while condition (\ref{eq:kriging unbiaseness condition}) implies ${\sum_{i=1}^N{\lambda_i}=1}$.
%Minimizing (\ref{eq:kriging variance}) yields 
%\begin{displaymath}\label{eq:kriging equation 1}\sum_{i=1}^N{\lambda_i\, \gamma(\mathbf{x}_i,\mathbf{x}_j)+\mu=\gamma(\mathbf{x}_i,\mathbf{x}_0)}\end{displaymath}
%  where $\mu$ is a Lagrange multiplier, while condition (\ref{eq:kriging unbiaseness condition}) implies
%\begin{displaymath}\label{eq:OK sum of weights}{\sum_{i=1}^N{\lambda_i}=1}\end{displaymath}
The resulting system of $N+1$ equations in $N+1$ unknowns ${\lambda_i}$ is known as the \emph{ordinary Kriging equations}. It is often expressed in matrix form as $\mathbf{A}\mathbf{\lambda}=\mathbf{b}$ with
\begin{equation}\label{eq:A matrix} \mathbf{A} = 
 \left[ 
 \begin{array}{ccccc}
  \gamma(\mathbf{x}_1,\mathbf{x}_1) & \gamma(\mathbf{x}_1,\mathbf{x}_2) & \cdots\, \gamma(\mathbf{x}_1,\mathbf{x}_N) & 1\\
  \gamma(\mathbf{x}_2,\mathbf{x}_1) & \gamma(\mathbf{x}_2,\mathbf{x}_2) & \cdots\, \gamma(\mathbf{x}_2,\mathbf{x}_N) & 1\\
  \vdots  & \vdots  & \vdots & \vdots \\
  \gamma(\mathbf{x}_N,\mathbf{x}_1) & \gamma(\mathbf{x}_N,\mathbf{x}_2) & \cdots\, \gamma(\mathbf{x}_N,\mathbf{x}_N) & 1\\
  1 & 1 & \cdots \hspace*{5mm} 1 \hspace*{7mm} & 0\\
 \end{array} 
 \right]
\end{equation}

\begin{displaymath} 
 \mathbf{\lambda^T} = 
 \left[ 
 \begin{array}{ccccc}
  \lambda_1 & \lambda_2 & \cdots\, \lambda_N & \mu \\
 \end{array} 
 \right] \qquad \sum_{i=1}^N{\lambda_i}=1
\end{displaymath}
\begin{displaymath} 
 \mathbf{b^T} = 
 \left[ 
 \begin{array}{ccccc}
  \gamma(\mathbf{x}_1,\mathbf{x}_0) & \gamma(\mathbf{x}_2,\mathbf{x}_0) & \cdots\, \gamma(\mathbf{x}_N,\mathbf{x}_0) & 1\\
 \end{array} 
 \right]
\end{displaymath} The weights ${\lambda_i}$, along with the Lagrange multiplier $\mu$, are obtained by inversing the $\mathbf{A}$ matrix 
\begin{equation}\label{eq:OK matrix equation lambda}{\mathbf{\lambda}=\mathbf{A}^{-1}\mathbf{b}}\end{equation}

The main interpolation steps with ordinary Kriging can now be articulated:
\begin{enumerate}

\item{Construct an experimental variogram by computing the experimental semivariance $\hat\gamma(\mathbf{h})$ for a range of separation distances $\Arrowvert\mathbf{h}\Arrowvert$.}\\

% as illustrated in  Fig.~\ref{fig:star distances}.}

% Fig.~\ref{fig:experimental variograms} depicts two \mbox{experimental variograms} of the $(e_1,e_2)$ components of ellipticity, computed on one of the GREAT10 Star Challenge PSF fields.

\item{Fit the experimental variogram against an authorized variogram model. The mathematical expressions for the most common authorized \mbox{\emph{theoretical variogram}} models are summarized in Table~\ref{table:variogram models}. After completion of this step, the $\gamma(\mathbf{x}_i,\mathbf{x}_j)$ value at any separation vector $\mathbf{h}=\mathbf{x}_j-\mathbf{x}_i$ can be calculated and used to compute the $\mathbf{A}$ matrix (\ref{eq:A matrix}).} \\

\item{Calculate interpolated values: derive the kriging weights ${\lambda_i}$ for each point of interest $\mathbf{x}_0$ by solving Eq.~(\ref{eq:OK matrix equation lambda}) and obtain the kriging estimate at $\mathbf{x}_0$ by substituting in (\ref{eq:generic interpolation formula 2}).}
\end{enumerate} 

\begin{table}
\renewcommand{\arraystretch}{1.25}
\caption{Authorized Kriging theoretical variogram models. The models in this table correspond to purely isotropic Kriging. More elaborate formulas exist for correcting geometrical anisotropy through the rescaling or rotation of coordinate axes along the direction of major spatial continuity.}             
\label{table:variogram models}      % is used to refer this table in the text
%\centering                                      % used for centering table
\begin{tabular}{l l l}          % centered columns (4 columns)
\hline\hline                        % inserts double horizontal lines
 Model & Expression\\    % table heading
\hline                                   % inserts single horizontalAfter obtaining line
    Pure Nugget & $\begin{array}{l l l} \gamma(h)=0 &  \qquad \qquad \qquad \qquad \hspace*{3.0mm} h = 0 \\
                                       \gamma(h)=c_0 \qquad c_0 \ge 0 & \qquad \qquad \qquad \qquad \hspace*{3.0mm} h > 0 \end{array}$ \\    % inserting body of the table
\hline                                   % inserts single horizontal line
    Spherical & $\begin{array}{l l l} \gamma(h)=c_0 + c\big\{\frac{3h}{2a}-\frac{1}{2}\big({\frac{h}{a}}\big)^3 \big\} & \qquad \qquad \qquad \quad h \leq a \\ 
                                     \gamma(h)=c_0 + c & \qquad \qquad \qquad \quad h > a \end{array}$  \\     % inserting body of the table
\hline                                   % inserts single horizontal line
    Exponential & $ \begin{array}{l l}\gamma(h)=c_0 + c\,\big\{1-exp\big(-\frac{h}{a}\big)\big\}\end{array}$ \\  % inserting body of the table
\hline                                   % inserts single horizontal line
    Gaussian & $ \begin{array}{l l}\gamma(h)=c_0 + c\,\big\{1-exp\big(-\frac{h^2}{a^2}\big)\big\}\end{array}$ \\  % inserting body of the table
\hline                                             %inserts single line
    Power & $ \begin{array}{l l l}\gamma(h)=c_0 + b\,{h\,}^p & \qquad \qquad\qquad b \ge 0,\, 0 \le p < 2 \end{array}$  \\% inserting body of the table
\hline                                             %inserts single line
\end{tabular}
\tablefoot{In the above expressions, $c_0=\lim_{h\to0}\gamma(h)$ is the so-called \emph{nugget} constant that represents measurement errors or indicates a spatially discontinuous process. The quantities $(c_0 + c)$ and $a$ respectively represent the variogram \emph{sill} and \emph{range}. The \emph{Pure Nugget} model corresponds to absence of spatial correlation.}
\end{table}
Most of the strengths of Kriging interpolation stem from the use of semivariance instead of pure geometrical distances. This feature allows Kriging to remain efficient in condition of sparse data and to be less affected by clustering and screening effects than other methods.\newline\indent In addition, as a true stochastic method, Kriging interpolation provides a way of directly quantifying the uncertainty in its predictions in the form of the Kriging variance specified in Eq.~(\ref{eq:kriging variance}).\newline

The sophistication of Kriging, on the other hand, may also be considered as one of its disadvantages. A thorough preliminary analysis of the data is required or at least strongly recommended prior to applying the technique \citep[e.g.][]{Tukey1977}. This can prove complex and time consuming.\newline\indent 
One should also bear in mind that Kriging is more computationally intensive than the other local interpolation methods described in this article.
The strong and weaker points of Kriging interpolation are highlighted in Table~\ref{table:kriging}.

%[Conlusions: sophisticated and costly, but this is very often worth it because it allows kriging to offer superior performance]

\begin{table}
\renewcommand{\arraystretch}{1.25}
\caption{Kriging interpolation: Pros and cons}              % title of Table
\label{table:kriging}      % is used to refer this table in the text
\begin{tabular}{l p{7.5cm}}          % centered columns (4 columns)
\hline\hline                        % inserts double horizontal lines
\multicolumn{2}{l}{Kriging} \\    % table heading
\hline                                   % inserts single horizontal line
\multirow{6}{*}{Pros} & Predictions based on a spatial statistical analysis of the data \\
                      & Best linear unbiased estimator (BLUE) \\
                      & Many forms of Kriging available, applicable to various data configurations \\
                      & Automatically accounts for clustering and screening effects; remains efficient in conditions of sparse data \\ 
                      & Can take into account variation bias toward specific directions (anisotropy) \\
                      & Able to quantify interpolation errors (Kriging variance) \\
\hline 
\multirow{4}{*}{Cons} & Overall complexity \\
                      & Requires care when modeling spatial correlation structures \\
                      & Assumptions of intrinsic stationarity may not be valid (drift) and be handled though an appropriate Kriging variant \\
                      & Most Kriging variants are exact (no smoothing) \\
                      & Kriging is more computationally intensive than other local methods \\
\hline                                             %inserts single line
\end{tabular}
\end{table}

\section{Applying spatial interpolation schemes on the GREAT10 Star Challenge data}
\label{section:applying on GREAT10 data}

In 2011, we participated in the GREAT10 Star Challenge competition \citep{GREAT10Handbook2010, GREAT10StarChallengeResults2012}, which allowed us to evaluate the performance of the interpolation schemes described above: those based on splines, inverse distance weighting (IDW), radial basis functions (RBF) and ordinary Kriging. To our knowledge, the only reference to a similar work in the field of astronomy is that of  \citet{Berge2012}. 

\begin{figure}
\resizebox{\hsize}{!}{\includegraphics[trim=4mm 5mm 5mm 4mm, clip]{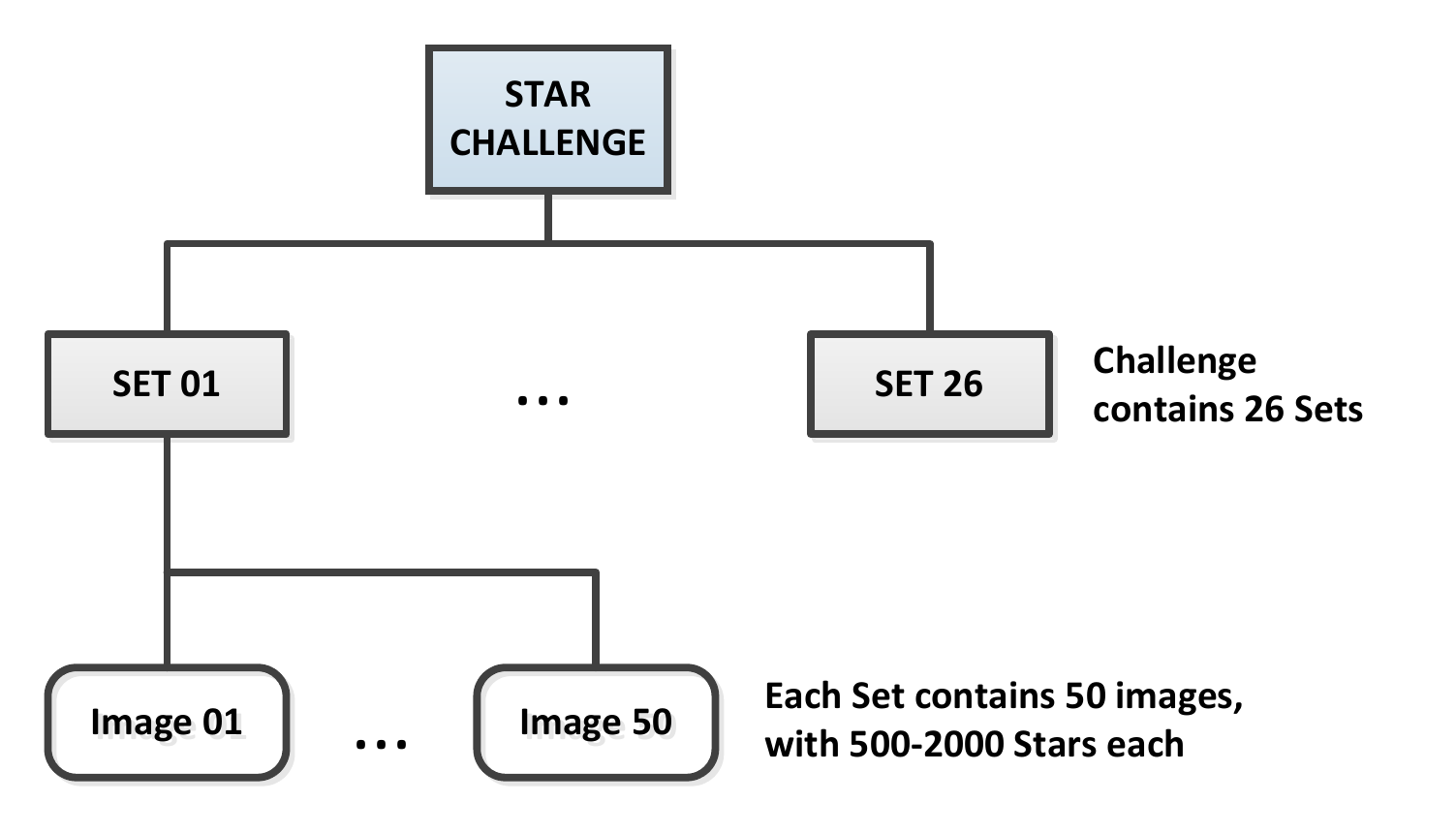}}
\caption{Star Challenge simulated data}
\label{fig:Star Challenge data}
\end{figure}

The GREAT10 Star Challenge ran from December 2010 to September 2011 as an open, blind competition. As illustrated in Fig.~\ref{fig:Star Challenge data},  the data consisted in $26$ datasets of $50$ PSF fields, each field containing between $500$ and $2000$ simulated star images and featuring specific patterns of variation. The stars images were supplied as non-overlapping, randomly-scattered $48\times48$ pixels postage stamps, altered by Gaussian noise.\newline\indent After completion of the challenge, it was revealed the stars had either a Moffat \citep{Moffat1969} or pseudo-Airy \citep{BornWolf1999, Kuijken2008} profile, with a telescope component model from \citet{Jarvis2008}. Depending on the sets, specific additional effects, such as Kolmogorov turbulence, were also incorporated.\newline\indent The challenge itself was to predict the PSF at $1000$ requested positions in each of the $1300$ PSF \mbox{fields (see Fig.~\ref{fig:PSF interpolation})}. 

%The Star Challenge required competitors to provide working solutions for modeling and predicting a noisy PSF field. We begin this section by presenting our own strategy for winning the challenge, along with a description of our PSF prediction pipeline. We continue with a critical review of our results in the competition and conclude by highlighting the respective merits of individual interpolation schemes.

\subsection{Which model for the PSF?}
\label{subsection:PSF model}
The first important step to make was to choose an appropriate model for the PSF. Indeed, before selecting a particular PSF interpolator, one has to decide on which type of data that interpolator will operate.

Essentially three PSF modeling approaches have been explored in the literature:
\begin{enumerate}
\item PSF as a combination of basis functions
\item PSF left in pixel form
\item PSF expressed in functional form
\end{enumerate}

To help choosing the right model for the data at hand, useful guidance is provided by the notions of complexity and sparsity, recently put forward by \citep{PaulinHenrikssonAmaraVoigt2008, PaulinHenrikssonRefregierAmara2009}. The complexity of a model is characterized by the amount of information required to represent the underlying PSF image, which can be expressed as the number of degrees of freedom (DoF) present in the model. The more sophisticated the model the greater the number of its DoF. Sparsity, on the other hand, is meant to describe how efficiently a model can represent the actual PSF with a limited number of DoF, that is, with a simple model. 
%A low complexity model will require less observations but have large residuals; inversely, a model with a high number of DoF will yield smaller residuals but large scatter on the fitted parameters, requiring more observable data. \newline

The  simulated star images looked relatively simple and we decided that the right level of sparsity could be achieved with PSF in functional form (the third option). We then assumed that the most likely PSF profile used to create the stars was either Airy or Moffat. We opted for an elliptically symmetric Moffat function for its simplicity and because the stars did not show significant diffraction spikes. Each star was thus assumed to have a light intensity \mbox{distribution} of the form:
\begin{displaymath}\label{eq:eliptical Moffat function} I(\xi)=I_0\,\big[1+\big({\frac{\xi}{\alpha}}\big)^2 \big]^{-\beta},\quad \xi=\sqrt{(x'-x_c)^2+\frac{(y'-y_c)^2}{q^2}} \end{displaymath}
In the above expression, $I_0$ is the flux intensity at $\xi=0$, $\xi$ \mbox{being} the radius distance from the centroid $(x_c,y_c)$ of the PSF to a spatial coordinate

\begin{equation}
 \left[ \begin{array}{c} x'-x_c\\ y'-y_c\\ \end{array} \right] = 
 \left[ \begin{array}{rl} \cos{\phi} & \sin{\phi} \\ -\sin{\phi} & \cos{\phi} \\  \end{array} \right]\,
 \left[ \begin{array}{c} x-x_c \\ y-y_c \\ \end{array} \right],
\end{equation}
obtained after \mbox{counterclockwise} rotation through an angle $\phi$ with respect to the $(0,\,x)$ axis. The quantity $\alpha=\mathit{FWHM}\, { [2^{1/\beta}-1] }^{-1/2}$ is the Moffat scale factor expressed in terms of the full width at half maximum ($\mathit{FWHM}$) of the PSF and the Moffat shape parameter $\beta$. Lastly, $q$ is the ratio of the semi-minor axis $b$ to the semi-major axis $a$ of the isophote ellipse, given by $q=b/a=(1-\arrowvert\mathbf{e}\arrowvert)/(1+\arrowvert\mathbf{e}\arrowvert)$, with $\arrowvert\mathbf{e}\arrowvert=\sqrt{{e_1}^2+{e_2}^2}$, $e_1=\arrowvert\mathbf{e}\arrowvert\,\cos 2\phi$ and $\,e_2=\arrowvert\mathbf{e}\arrowvert\,\sin 2\phi$. 

\begin{figure}[t]
\resizebox{\hsize}{!}{\includegraphics[trim=5mm 5mm 5mm 5mm, clip]{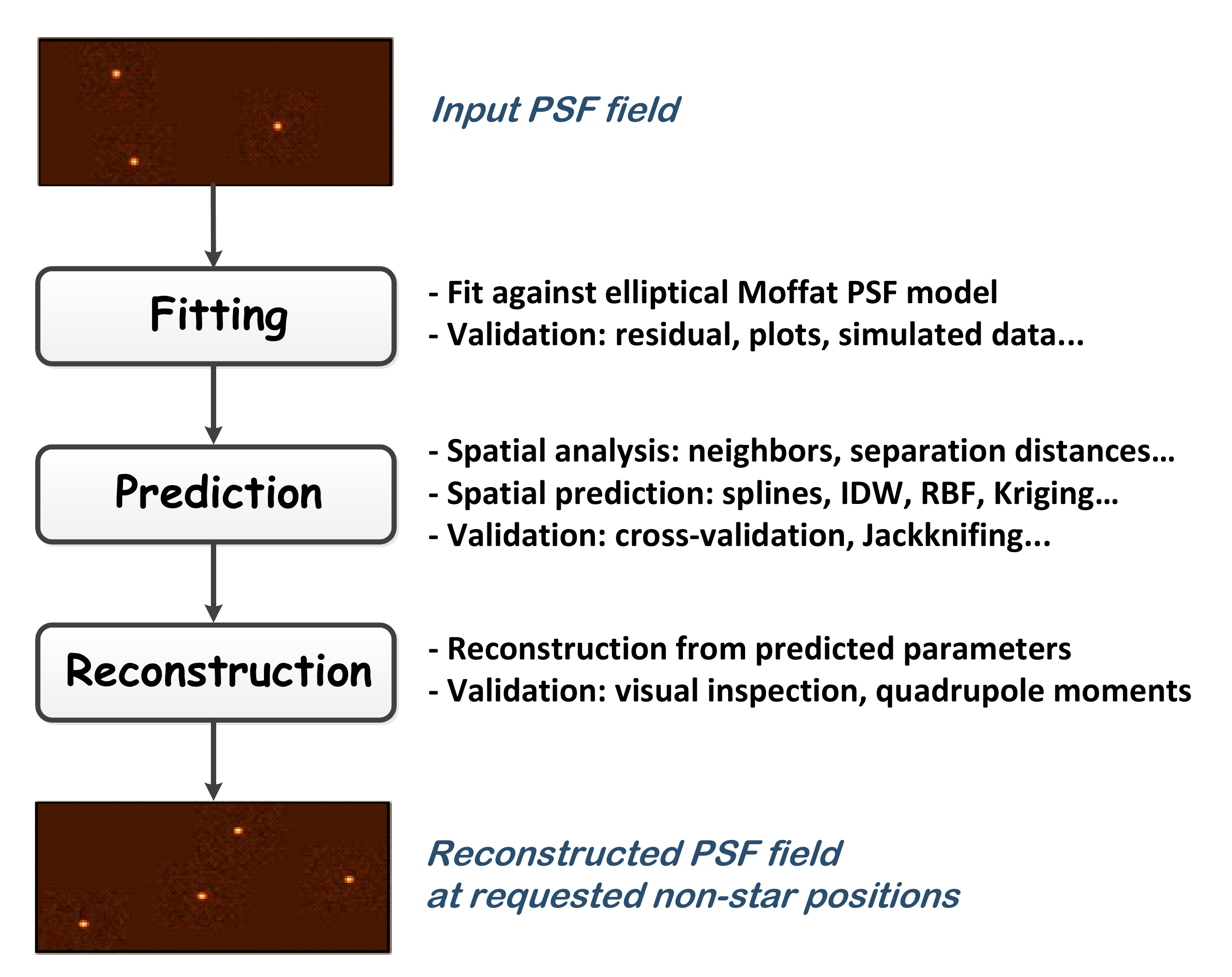}}
\caption{The three-stage PSF prediction pipeline we used to compete in the Star Challenge. Elliptical Moffat profiles are fitted to the stars contained in the input Star Challenge PSF field; the model resulting parameters are then individually interpolated across the field at requested locations, using one of our PSF spatial interpolator. Lastly, the star images are reconstructed from the set of Moffat parameters predicted in the previous stage.}
\label{fig:PSF prediction pipeline}
\end{figure}

\subsection{Our PSF prediction pipeline}
\label{subsection:prediction pipeline}

The three-stage PSF prediction pipeline we used in the Star Challenge is sketched in Fig.~\ref{fig:PSF prediction pipeline}. The purpose of the \emph{Fitting} stage is to produce a catalog of estimated full width at half maximum (FWHM) and ellipticity values of the stars found at known spatial positions within the input Star Challenge PSF image.

In the \emph{Prediction}  stage, that catalog is processed by an interpolation algorithm and a catalog is produced with estimated FWHM and ellipticities at new positions in the same image. Competitors were required to submit their results in the form of FITS Cube images \citep{GREAT10Handbook2010}. In the \emph{Reconstruction} stage, each star in a PSF field is thus reconstructed using that format from the interpolated quantities predicted in the \emph{Prediction}  stage.  A more detailed description of the pipeline is given in Appendix~\ref{section:appendix PSF prediction pipeline}.
\subsection{Cross-validation and Jackknifing}
\label{subsection:cross-validation}
The Star Challenge was a \emph{blind} competition. The true answers being unknown, it was essential to find ways to evaluate how far the actual results were from the truth. To assess the fitting accuracy in the first stage of the pipeline we could rely somewhat on the analysis of the residuals between observed and fitted star images. But when it came to evaluate prediction results, we had no such residuals to help us appraise the accuracy of the \mbox{interpolation} algorithm: we could only rely on the fitted observations $Z(\mathbf{x}_i)$.
The use of cross-validation and Jackknifing provided a satisfactory solution to this problem. 

\subsubsection*{Cross-validation}
Cross-validation (CV) is a resampling technique frequently used in the fields of machine learning and data mining to evaluate and compare the performance of predictive models \citep{Stone1974, Geisser1975, IsaaksSrivastava1989, Browne2000}.

%In its simplest form, \mbox{\emph{Leave-One-Out} cross-validation} (LOOCV), one value from a dataset of $n$ observations $Z(\mathbf{x}_i)$ is excluded at a location $\mathbf{x}_k$ (which we note $Z{(\mathbf{x}_{[k]})}$) and an estimate $\hat Z{(\mathbf{x}_{[k]})}$ at $\mathbf{x}_k$ is made using the remaining $n-1$ observations using the prediction algorithm. This process is repeated in turn for each $Z(\mathbf{x}_i)$ in the dataset. Since the values at excluded locations are known, various diagnostic statistics, such as those listed in Table~\ref{table:diagnostic_statistics}, can then be applied to compare the $Z(\mathbf{x}_i)$ and $\hat Z{(\mathbf{x}_{[i]})}$.

In the context of the Star Challenge, we used CV to both evaluate the performance of an interpolation method and tune the free parameters of the underlying interpolation models. 

As explained earlier, the deterministic interpolation methods (IDW, RBF, splines) we tested in the competition did not provide any quantification of residual errors. The first three diagnostic statistics mentioned in Table~\ref{table:diagnostic_statistics} provided a good indication of the level of accuracy reached. This technique was useful for Kriging as well because we could directly compare the mean error (ME) and mean squared error (MSE) provided by CV: Kriging being an unbiased estimator, we expected ME to be nearly zero, the MSE to be close to the Kriging variance provided by Eq.~(\ref{eq:kriging variance}) and the mean squared deviation ratio (MSDR) to be around unity. 

CV also proved useful for tuning the free parameters of the models behind the interpolation schemes, as mentioned in Appendix~\ref{subsection:pipeline implementation}. For instance, for RBF interpolation, we could rapidly try and discard the cubic, quintic, Gaussian and inverse multiquadric kernel functions. Another example was the ability to find the best search neighborhood size for local distance-based interpolation methods.

\subsubsection*{Jackknifing}

The Jackknifing resampling technique was first proposed by \citet{Quenouille1956} and further developed by \citet{Tukey1958}. A classical review on that subject is that of \citet{Miller1974}.  See also \citet{Efron1982, EfronGong1983, Davis1987, Tomczak1998} for more general discussions on the use of CV in connection to Jackknifing.

%With Jackknifing, the set of locations where new estimates are made is distinct from the locations where input observations are available, whereas, with cross-validation, both sets were the same. In other words, with Jackknifing, the prediction algorithm is forced to process data at locations it had never seen before.

To Jackknife a Star Challenge PSF field image, we would typically split the set of input coordinates into two equally-sized sets of star locations,  i.e. $1000$ randomly-selected star centroid positions from a set of $2000$, one used for input and one used for prediction. We would then interpolate the PSF of the prediction set based on the PSF of the input set.

%Further Texts on cross-validation, Jackknifing and related statistical techniques can be found for example in \citet{Efron1982, EfronGong1983, Davis1987, Tomczak1998}.

\begin{table}[t]
\renewcommand{\arraystretch}{1.25}
\caption{Common diagnostic statistics for use with cross-validation and Jackknifing.}             % title of Table
\label{table:diagnostic_statistics}      % is used to refer this table in the text
%\centering                                      % used for centering table
\hfill
\begin{tabular}{p{3.0cm} l}          % centered columns (4 columns)
\hline\hline                        % inserts double horizontal lines
Statistics & Expression\\    % table heading
\hline                                % inserts single horizontal line    
    Mean error & $ME=\frac{1}{n}\sum_{i=1}^n{\big[Z(\mathbf{x}_i) - Z{(\mathbf{x}_{[i]})\big]}}$   \\[1mm]    % inserting body of the table
    Mean squared error & $MSE=\frac{1}{n}\sum_{i=1}^n{\big[Z(\mathbf{x}_i) - Z{(\mathbf{x}_{[i]})}\big]^2}$ \\[1mm]
    Mean absolute error & $MAE=\frac{1}{n}\sum_{i=1}^n{\big\Arrowvert Z(\mathbf{x}_i) - Z{(\mathbf{x}_{[i]})}\big\Arrowvert}$ \\[1mm]
%    Mean square deviation ratio & $MSDR=\frac{1}{n}\sum_{i=1}^n\frac{\big[Z(\mathbf{x}_i) - Z{(\mathbf{x}_{[i]})}\big]^2}{\sigma_{[i]}} $ \\[1mm]

    Mean squared deviation ratio & $MSDR=\frac{1}{n} \sum_{i=1}^n\big\{{\big[Z(\mathbf{x}_i) - Z{(\mathbf{x}_{[i]})}\big]^2} / \sigma_{[i]}^2\big\} $ \\[1mm]
\hline                                             %inserts single line
\end{tabular}
\end{table}

%\begin{figure}[t]
%\resizebox{\hsize}{!}{\includegraphics[trim=3mm 3mm 3mm 3mm, clip]{fig_star_challenge_leaderboad_red.pdf}}
%\caption{GREAT10 Star Challenge final leaderboard, with the submissions corresponding to the methods described in this paper. Our results are shown in red.}
%\label{fig:Star Challenge leaderboard}
%\end{figure}

\section{Analyzing our GREAT10 Star Challenge results}
\label{section:results analysis}

\subsection{Results on the Star Challenge data}
\label{subsection:official results}

\begin{table}[t]
\renewcommand{\arraystretch}{1.25}
\caption{Final results obtained by the B-SPLINE, IDW, Kriging and RBF methods in the Star Challenge, sorted by decreasing P-factors. The B-splines method obtained the highest P-factor of the competition while the remaining four achieved the next highest scores.}
\label{table:Star Challenge leaderboard}      % is used to refer this table in the text
%\centering                                      % used for centering tableto avoid sharp peaks or troughs in the output surface
\begin{tabular}{clll}          % centered columns (4 columns)
\hline                                             %inserts single line{10^{-7}
\hline                       % inserts double horizontal lines
Rank & PSF Interpolation method & P & $\sigma_{sys}^2$ \\   % table heading
\hline 
   1 & Basis Spline (B-splines) & $13.29$ & $7.53\times10^{-5}$ \\
   2 & inverse distance weighting (IDW) & $13.17$ & $7.59\times10^{-5}$ \\
   3 & radial basis function (RBF) & $12.72$ & $7.86\times10^{-5}$ \\
   4 & radial basis function (RBF thin) & $12.61$ & $7.93\times10^{-5}$ \\
   5 & ordinary Kriging (OK) & $7.23$ & $1.38\times10^{-4}$ \\
\hline                                            
\end{tabular}
\end{table}

\begin{table}[t]
\renewcommand{\arraystretch}{1.25}
\caption{Average values of the performance metrics $E$ and $\sigma$ (see Sect.~\ref{subsection:performance metrics}) over all sets, obtained by the B-SPLINE, IDW, Kriging and RBF methods in the Star Challenge.}
\label{table:Star Challenge statistics}      % is used to refer this table in the text
%\centering                                      % used for centering tableto avoid sharp peaks or troughs in the output surface
\begin{tabular}{lllll}          % centered columns (4 columns)
\hline                                             %inserts single line{10^{-7}
\hline                       % inserts double horizontal lines
Method & $E(e)$  & $\sigma(e)$ & $E(R^2)$ & $\sigma(R^2)$\\
\hline 
 B-splines & $2.03\times10^{-2}$ & $8.57\times10^{-4}$ & $1.90\times10^{-1}$ & $8.25\times10^{-4}$ \\
 IDW & $2.04\times10^{-2}$ & $8.69\times10^{-4}$   & $1.92\times10^{-1}$ & $1.07\times10^{-3}$ \\
 RBF & $2.26\times10^{-2}$ & $9.73\times10^{-4}$  & $1.98\times10^{-1}$ & $1.39\times10^{-3}$  \\
 Kriging & $3.17\times10^{-2}$ & $1.26\times10^{-3}$  & $2.22\times10^{-1}$ & $2.18\times10^{-3}$  \\
\hline                                             %inserts single line
\end{tabular}
\end{table}

The results obtained in the Star Challenge by the B-splines, IDW, Kriging, RBF and RBF-thin PSF interpolation schemes are shown in Table~\ref{table:Star Challenge leaderboard}. 

The B-splines method won the Star Challenge while the remaining four achieved the next highest scores of the competition. %As mentioned in Appendix~\ref{subsection:pipeline implementation}, the RBF and RBF-thin methods, despite sharing the same interpolation algorithm, respectively use the linear and thin-plate kernel functions (see Table~\ref{table:popular RBF kernels}).

The quantity $P$ refers to the so-called \emph{P-factor}, specified in \citet{GREAT10Handbook2010, GREAT10StarChallengeResults2012}. That P-factor is defined so as to measure the average variance over all images between the estimated and true values of two key PSF attributes: its size $R$ and ellipticity modulus $e=\arrowvert\mathbf{e}\arrowvert$, estimated using second brightness moments computed over the reconstructed PSF images. Since the GREAT10 simulated star images have either Moffat or Airy profiles, $R$ is actually an estimator of the FWHM of the stars.

The $\sigma_{sys}^2$ quantity is related to the P-factor by $\sigma_{sys}^2 = 10^{-3} / P$ and represents a \mbox{total} residual variance in the measurement of the PSF. It approximates the corresponding metric specified in \citet{AmaraRefregier2008,PaulinHenrikssonAmaraVoigt2008, PaulinHenrikssonRefregierAmara2009}.

\subsection{Performance metrics}
\label{subsection:performance metrics}

In this article, we do not rely on the P-factor as a metric for \mbox{assessing} the performance of our methods, for the following reasons. 
Firstly, the P-factor is specific to the Star Challenge and is not mentioned anywhere else in the literature on PSF interpolation. Secondly, we are really interested in knowing the individual accuracy of ellipticity and size but $P$ only appraises the combined performance of these quantities.

To assess the performance of an interpolator, we calculate instead the root mean squared error (RMSE) and standard \mbox{error} on the mean (SEM) of the residuals between true and calculated values of PSF ellipticity and size. As in \citet{PaulinHenrikssonAmaraVoigt2008, GREAT10StarChallengeResults2012},  we adopt the ellipticity modulus  $e={(e_1^2 + e_2^2)}^{1/2}$ and size squared $R^2$ as respective measures of ellipticity and size, and define the corresponding residuals as
\begin{displaymath}
\delta(e)=e_{calc} - e_{true}, \qquad \delta(R^2)=R_{calc}^2 - R_{true}^2
\end{displaymath}
As regards PSF ellipticity, we adopt as  performance metrics
\begin{displaymath}
 E(e)=\mathit{RMSE}(\delta(e) / 2), \quad \sigma(e)=\mathit{SEM}(\delta(e) / 2)
 \end{displaymath}
while for  PSF size, we  evaluate
 \begin{displaymath}
 E(R^2)=\mathit{RMSE}(\delta(R^2)) / \langle R_{true}^2 \rangle, \quad  \sigma(R^2)=\mathit{SEM}(\delta(R^2)) / \langle R_{true}^2 \rangle
 \end{displaymath} where the angle brackets $\langle$ and  $\rangle$ denote averaging.
The factor $2$ in the expressions of $E(e)$ and $ \sigma(e)$ arises because ellipticity has two components.
We calculate these metrics over the $N=1000$ stars in each of the $50$ images of each set.

The quantity $E$ provides a measure of the global accuracy of the interpolator (bias and precision combined) while $\sigma$ provides insights into the variance of the residuals. The exact expressions for these performance metrics are given in Table~\ref{table:performance metrics}. 

\begin{table}[t]

\renewcommand{\arraystretch}{1.75}
\caption{Performance metrics used in this article. The angle brackets $\langle$ and  $\rangle$ denote averages and ``$\mathit{stdev}$''  the standard deviation. These statistics are calculated over the $N=1000$ stars in each of the $50$ images of each set.} 
\label{table:performance metrics}      
%\centering   
\begin{tabular}{ll}          
\hline\hline                       
 PSF attribute & Metrics\\    
\hline        
    PSF Ellipticity & $\begin{array}{ll}  E(e)= \sqrt {\big \langle { (e_{calc} - e_{true}} ) ^2\big\rangle} \,  / \, 2 \\
                                       \sigma(e)=\mathit{stdev}\,(e_{calc} - e_{true} ) \,  / \sqrt{2} \, / \sqrt{N} \end{array}$ \\                 
 \hline                                            

    PSF Size & $\begin{array}{ll}  E(R^2)= \sqrt {\big {\langle (R^2_{calc} - R^2_{true})^2 \big \rangle }} \, / \, \langle {R^2_{true}} \rangle \\
                                       \sigma(R^2)=\mathit{stdev}\,({R^2_{calc}} - {R^2_{true}}) \, / \, \langle {R^2_{true}} \rangle \, / \sqrt{N} \end{array}$\\                             
\hline                                           
\end{tabular}
\end{table}

\subsection{Analysis of the Star Challenge results}
\label{subsection:general observations}
%\begin{itemize}
%\item The results from IDW, RBF, splines are close. Kriging lags behind.
%\item Q is low: error can arise from three main sources: fitting, interpolation and reconstruction
%\item 
%\end{itemize}

The performance metrics of B-splines, IDW, RBF and Kriging are given in Table~\ref{table:Star Challenge statistics}. The results of RBF and RBF-thin being very close, we no longer distinguish these two interpolators in the reminder of this paper and only mention them collectively as RBF.

Since a detailed analysis of the Star Challenge results of B-splines, IDW, RBF and Kriging as already been performed in \citet{GREAT10StarChallengeResults2012}, a similar analysis would be redundant here.
We do have, however, a couple of observations to make, based on the metrics in Tables~\ref{table:Star Challenge leaderboard} and \ref{table:Star Challenge statistics}.\\ 

%Before conducting a more detailed analysis we make a few \mbox{general observations} from the leaderboard Fig.~\ref{fig:Star Challenge leaderboard}.

We observe that the global $\sigma_{sys}^2$ variance of the most successful interpolation method is of the order of $10^{-4}$. As demonstrated in \citet{AmaraRefregier2008,PaulinHenrikssonAmaraVoigt2008} and confirmed by \citet{KitchingAmara2009}, future large surveys will need to constrain the total variance in the systematic errors to \mbox{$\sigma_{sys}^2<{10^{-7}}$}, which corresponds to $E(e) \lesssim{10^{-3}}$ and $E(R^2)\lesssim {10^{-3}}$. The Star Challenge results thus tend to suggest that a $\sim10$ improvement in $E(e)$ and a $\sim100$ improvement in ${E(R^2)}$ are still required for achieving that goal.

Secondly, since we have been using a three-stage pipeline as described in Sect.~\ref{subsection:prediction pipeline}, each stage, fitting, interpolation and reconstruction, can potentially contribute to the final error in size and ellipticity. 
Investigations following the publication of the true size and  \mbox{ellipticity} values after the end of the Star Challenge, have led us to conclude fitting was actually the main  \mbox{performance} limiting factor, not the interpolation or reconstruction process. 

Also, the comparatively lower performance of Kriging is not related to the interpolation algorithm itself, but is actually  due to an inadequate fitting setup, that was subsequently fixed for \mbox{B-splines}, IDW and RBF submissions.\\

As the main goal of this article is to assess the respective merits of the interpolation methods, we wish to eliminate all inaccuracies related to fitting. To achieve this, we use instead of our fitted ellipticity and FWHM estimates at known positions, the \emph{true} input values, kindly supplied to us by the GREAT10 team. We interpolate these true input values at the expected target positions and then measure the error made by the interpolators. We present and analyze the corresponding results in the next section.

\section{Comparing PSF spatial interpolation schemes}
\label{section:comparative analysis}

The results presented in this section are based on true FWHM and ellipticity values at known positions in the Star Challenge PSF images. We are thus confident that error statistics we obtained truly reflect the performance of the PSF interpolation methods and are not influenced in any way by inaccuracies due to the fitting of our PSF model or to the image reconstruction processes.
% In other words, we have bypassed the \emph{Fitting}  and \emph{Reconstruction} stages of our PSF pipeline to only ran the  \emph{Prediction} pipeline.\newline

We compare below the respective performance of five PSF spatial interpolation schemes: 
\begin{itemize}
\item The four interpolation schemes introduced in Sect.~\ref{section:PSF interpolation schemes} that competed in the Star Challenge: B-splines, IDW, RBF and ordinary Kriging. \\
\item An additional scheme, labeled \emph{Polyfit}, which corresponds to a least-squares bivariate polynomial fit of the PSF, similar to that typically used in weak lensing studies (see Sect.~\ref{subsection:polynomial fitting}). 
% A fifth-degree polynomial gave the best predictions, higher degrees degrading the fit.
\end{itemize}

The metric values reflecting the average accuracy $E$ and error on the mean $\sigma$ for these five interpolation schemes are given in Table~\ref{table:new Star Challenge statistics}.

\subsection{Overall performance}
\label{subsection:overall accuracy}

The $E$ and $\sigma$ metrics on ellipticity and size after interpolation with all five methods are given in Table~\ref{table:new Star Challenge statistics}. These results lead to the following observations:

\begin{itemize}
\item If we compare Tables~\ref{table:new Star Challenge statistics} and \ref{table:performance metrics} we observe a $\sim$ 100-fold decrease of $E(R^2)$ for all interpolators. This confirms that the fitting of PSF sizes was the main limitation that prevented us from reaching better results in the Star Challenge. In comparison, the fitting of ellipticities was quite good.\\  

\item if we now concentrate on Table~\ref{table:new Star Challenge statistics}, we find that the RBF interpolation scheme based on the use of \mbox{radial basis functions}, has the highest accuracy and smallest error of the mean, both on size and ellipticity. We also observe that $E(e)\sim10^{-2}$ whereas $E(R^2)\sim10^{-3}$. This is because these statistics are averages over 26 image sets with different characteristics (see Sect.~\ref{subsection:influence of PSF features simulated in the image}). In reality, $E(e)$ varies \mbox{between $\sim10^{-2}$ and $\sim10^{-4}$}, whereas $E(R^2)\sim10^{-3}$ regardless of the sets. \\

\item If we consider $E(e)$ in particular, two groups emerge. The first one contains RBF, IDW and Kriging, with \mbox{$E(e)\lesssim1.8\times10^{-2}$}. The interpolators of the second group, B-splines and Polyfit with $E(e)\gtrsim2.3\times10^{-2}$. We will see below that this is essentially due to the better accuracy of local interpolators on turbulent sets as regards ellipticity.
If we focus on $E(R^2)$, the distinction between local and global interpolation schemes disappears. RBF and Polyfit stand out from the others with $E(R^2)\simeq5\times10^{-3}$. We also note that the accuracy of IDW on size is worse by several order of magnitude. \\
 
\item The errors on the mean $\sigma(e)$ and $\sigma(R^2)$ are of the order of $10^{-4}$ for all five schemes. As was observed for $E(e)$, we find that the local interpolators RBF, IDW and Kriging reach better $\sigma(e)$ values compared to global ones, B-splines and Polyfit. As for $\sigma(R^2)$, the best values are reached by RBF and Polyfit, similarly to what was found for $E(R^2)$.

\end{itemize}

\begin{table}[t]
\renewcommand{\arraystretch}{1.25}
\caption{Average values of the performance metrics $E$ and $\sigma$ (see Sect.~\ref{subsection:performance metrics}) over all sets, based on the true input ellipticities and sizes.}
\label{table:new Star Challenge statistics} 

\begin{tabular}{lllll}          
\hline                                           
\hline                       
Method & $E(e)$  & $\sigma(e)$ & $E(R^2)$ & $\sigma(R^2)$\\
\hline 
 RBF & $1.73\times10^{-2}$ & $7.18\times10^{-4}$  &  $4.58\times10^{-3}$ & $1.44\times10^{-4}$  \\
 IDW & $1.78\times10^{-2}$ & $7.24\times10^{-4}$   & $9.25\times10^{-3}$ & $2.91\times10^{-4}$ \\
 Kriging & $1.82\times10^{-2}$ & $7.09\times10^{-4}$  & $6.47\times10^{-3}$ & $2.03\times10^{-4}$  \\
 Polyfit & $2.29\times10^{-2}$ &  $7.52\times10^{-4}$ & $5.16\times10^{-3}$ & $1.62\times10^{-4}$ \\
 B-splines & $2.33\times10^{-2}$ & $7.39\times10^{-4}$  &  $6.45\times10^{-3}$ & $2.04\times10^{-4}$ \\
\hline                                            
\end{tabular}
\end{table}

\subsection{Influence of PSF features simulated in the images}
\label{subsection:influence of PSF features simulated in the image}

As explained in the Star Challenge result paper \citep{GREAT10StarChallengeResults2012}, the image sets were designed to simulate typical PSF features found in real astronomical images. Each set implements a unique combination of characteristics against which a method can be evaluated. All 50 images within a set share the same broad features, but differ in the way star positions, sizes and ellipticities are spatially distributed across the field.  

\begin{figure}[!]
\resizebox{\hsize}{!}{\includegraphics[trim=9mm 4mm 4mm 8mm, clip]{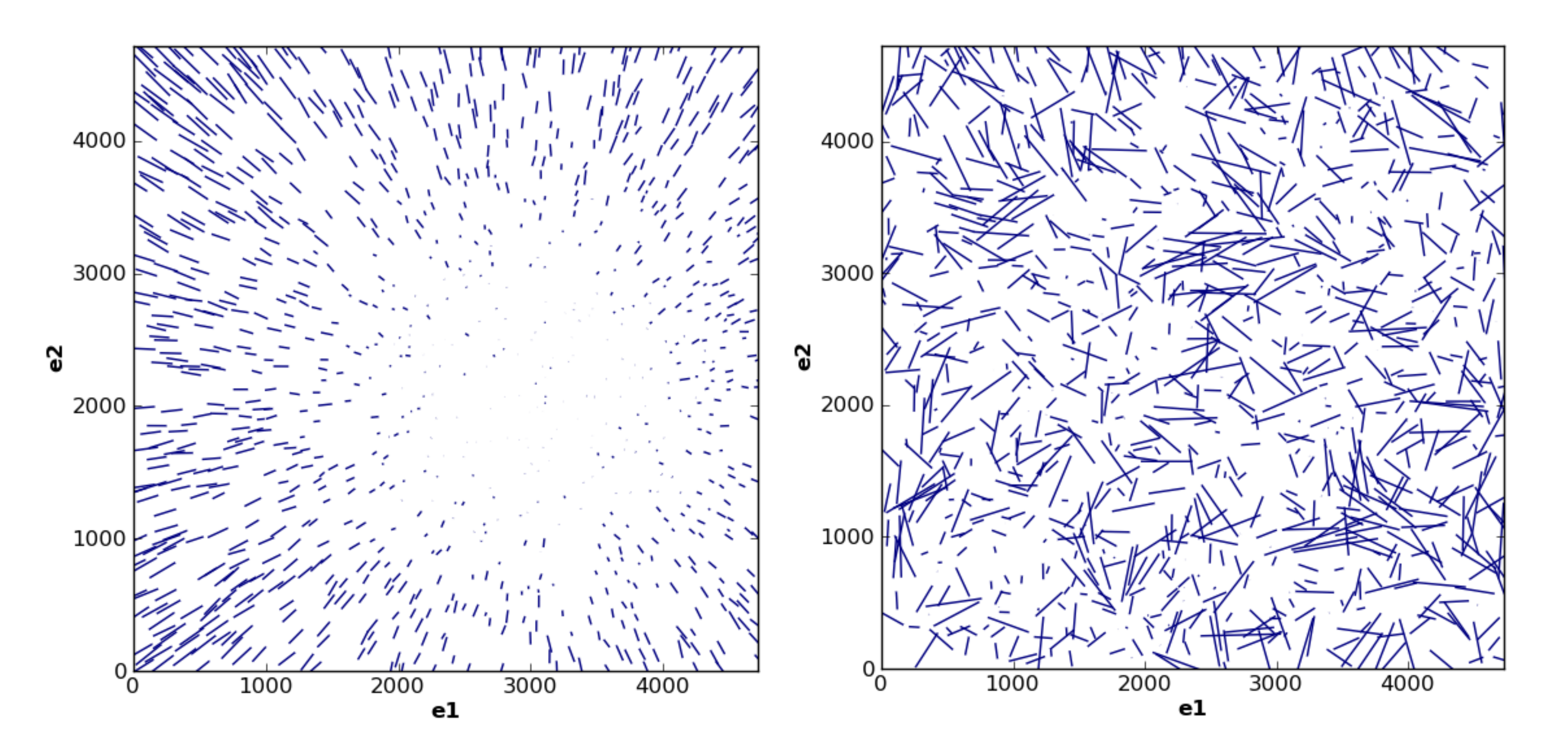}}
\caption{A Star Challenge non-turbulent PSF (left) compared with a turbulent PSF (right). Each ``whisker'' represents the amplitude $\arrowvert\mathbf{e}\arrowvert$ of the ellipticity of stars. The largest whisker in the left hand side image corresponds to an ellipticity of 0.16. The right hand side image has a \mbox{maximum} ellipticity of 0.37. The ellipticity plots have respectively been made from the first PSF field image of sets 8 and 14.}
\label{fig:turbulent_non-turbulent}
\end{figure}

\begin{figure*}[!]
\resizebox{\hsize}{!}{\includegraphics[trim=4mm 4mm 4mm 4mm, clip]{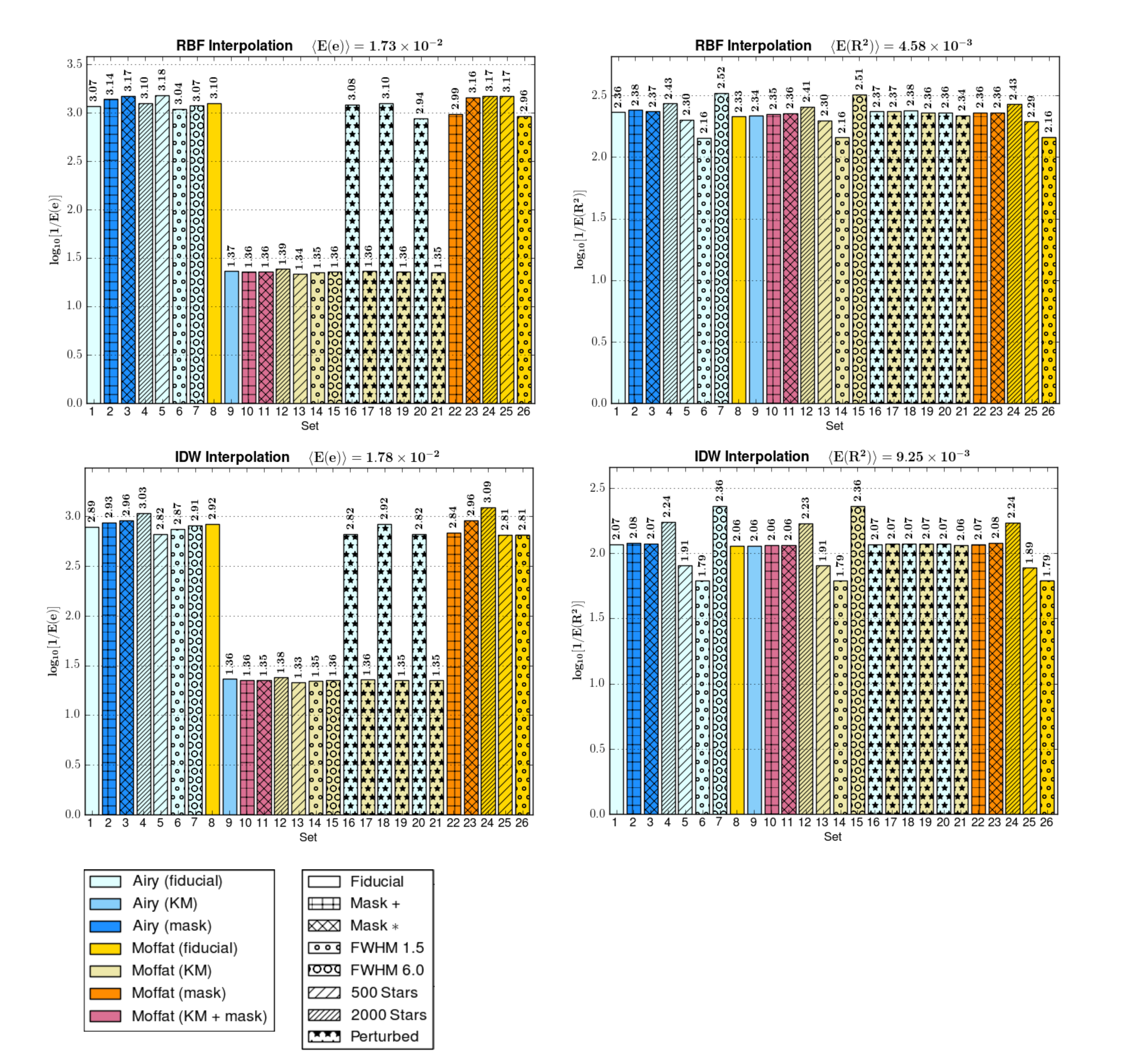}}
\caption{Accuracy per set for the RBF and IDW interpolation methods. Sets with pseudo-Airy and Moffat are respectively colored in different shades of blue and orange, as specified in the legend at the bottom left of the Figure. The various patterns contained in the left hand-side legend indicate the types of physical PSF features simulated in the images. The values on the bars correspond to $log_{10}({1/E(e)})$ and $log_{10}({1/E(R^2)})$ depending on the quantity plotted, so the taller the bar the greater the corresponding accuracy.}
\label{fig:plots metrics part 1}
\end{figure*}

\begin{figure*}[!]
\resizebox{\hsize}{!}{\includegraphics[trim=4mm 4mm 4mm 4mm, clip]{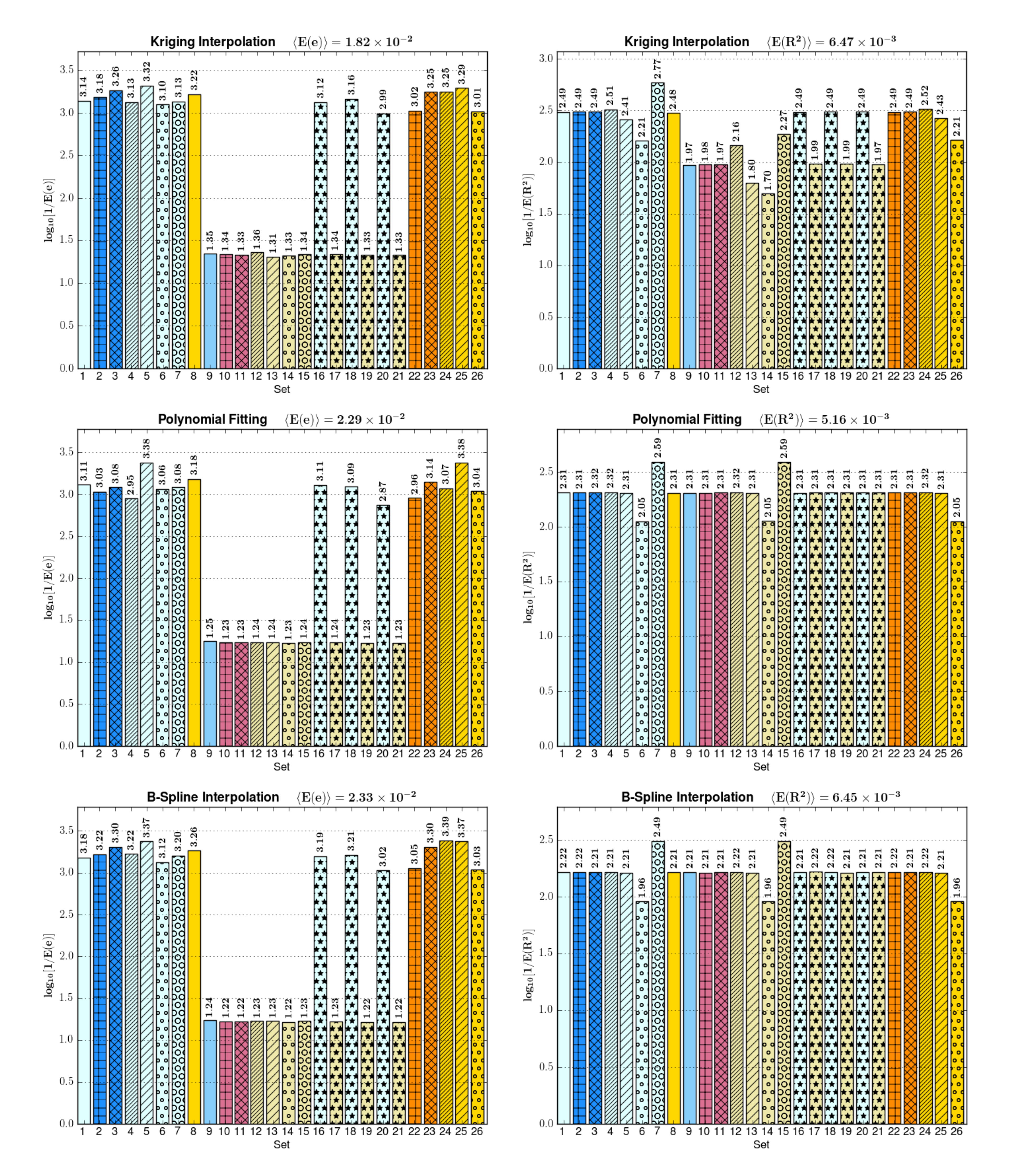}}
\caption{Accuracy per set for Kriging, polynomial fitting and B-splines. The legend is the same as that used in Fig~\ref{fig:plots metrics part 1}. The values on the bars correspond to $log_{10}({1/E(e)})$ and $log_{10}({1/E(R^2)})$ depending on the quantity plotted, so the taller the bar the greater the corresponding accuracy.}
\label{fig:plots metrics part 2}
\end{figure*}

The various PSF features tracked in the images are outlined below:

\begin{itemize}
\item \textbf{PSF model}: the fiducial PSF model includes a static and a dynamic component. The static component is based on a pseudo-Airy \citep{BornWolf1999, Kuijken2008} or Moffat \citep{Moffat1969} functional form, depending on the set. The dynamic component made the ellipticity and size of individual stars vary spatially across the image of the PSF field.\\

\item \textbf{Star size}: the images from most of the sets share the same ``fiducial'' 3-pixel FWHM, except sets 6, 14, 26 and sets 7, 15 whose images have respectively a FWHM of 1.5 and 6 pixels.\\

\item \textbf{Masking}: sets 2, 10, 22 have a 4-fold symmetric mask denoted as ``+'' and sets 3, 11, 23 have a 6-fold mask symbolized by a ``$\ast$''. Images from all other sets are unmasked.\\

\item \textbf{Number of stars}: the majority of images contain 1000 stars. Sets 4, 12, 24 are denser, with 2000 stars, whereas sets 5, 13, 25 are sparser, with only 1500 stars.\\

\item \textbf{Kolmogorov turbulence (KM)}: an attempt was made on sets 9 to 15, 17, 19 and 21 to simulate the effect of atmospheric turbulence by including a Kolmogorov spectrum in PSF ellipticity. See \citet{HeymansRowe2011, GREAT10StarChallengeResults2012} for the details. Fig.~\ref{fig:turbulent_non-turbulent} shows side by side a non-turbulent and a turbulent PSF.\\

\item \textbf{Telescope effect}: a deterministic component was included in sets 17, 19 and 21 to reproduce effects from the \mbox{telescope optics} on the PSF ellipticity and size, essentially primary astigmatism, primary defocus and coma \citep{BornWolf1999}, based on the model of  \citet{JarvisJain2004}.
\end{itemize}
In order to determine how interpolation schemes are affected by the aforementioned PSF characteristics, we have computed for each of them the performance metrics per individual image sets. We have plotted the metrics $E(e)$ and $E(R^2)$ in Figs.~\ref{fig:plots metrics part 1} and \ref{fig:plots metrics part 2}. We analyze the results below.

%In order to determine how interpolation schemes are affected by the aforementioned PSF characteristics, we have computed for each of them the performance metrics per individual image sets and provided the results in Appendix~\ref{section:appendix performance metrics per set}. We have also plotted the metrics $E(e)$ and $E(R^2)$ in Figs.~\ref{fig:plots metrics part 1} and \ref{fig:plots metrics part 2}. 

\begin{table}[!]
\renewcommand{\arraystretch}{1.25}
\caption{Non-turbulent sets: average values of $E$ and $\sigma$.}
\label{table:Star Challenge statistics non-turbulent}                                          
\begin{tabular}{lllll}         
\hline                                       
\hline                      
Method & $E(e)$  & $\sigma(e)$ & $E(R^2)$ & $\sigma(R^2)$\\
\hline 
 RBF & $8.26\times10^{-4}$ & $3.60\times10^{-5}$ & $4.59\times10^{-3}$ & $1.45\times10^{-4}$  \\
 IDW & $1.28\times10^{-3}$ & $5.67\times10^{-5}$ & $9.37\times10^{-3}$ & $2.95\times10^{-4}$ \\
 Kriging & $7.06\times10^{-4}$ & $3.16\times10^{-5}$ & $3.57\times10^{-3}$ & $1.13\times10^{-4}$  \\
 Polyfit & $8.37\times10^{-4}$ &  $3.73\times10^{-5}$ & $5.23\times10^{-3}$ & $1.64\times10^{-4}$ \\
 B-splines & $6.28\times10^{-4}$ & $2.80\times10^{-5}$  &  $6.53\times10^{-3}$ & $2.06\times10^{-4}$ \\
\hline                                            
\end{tabular}
\end{table}

\begin{table}[!]
\renewcommand{\arraystretch}{1.25}
\caption{Turbulent sets: average values of $E$ and $\sigma$.}
\label{table:Star Challenge statistics turbulent}                                          
\begin{tabular}{lllll}         
\hline                                       
\hline                      
Method & $E(e)$  & $\sigma(e)$ & $E(R^2)$ & $\sigma(R^2)$\\
\hline 
 RBF & $4.36\times10^{-2}$ & $1.81\times10^{-3}$ & $4.57\times10^{-3}$ & $1.44\times10^{-4}$  \\
 IDW & $4.42\times10^{-2}$ & $1.79\times10^{-3}$ & $9.05\times10^{-3}$ & $2.85\times10^{-4}$ \\
 Kriging & $4.61\times10^{-2}$ & $1.79\times10^{-3}$  & $1.11\times10^{-2}$ & $3.49\times10^{-4}$  \\
 Polyfit & $5.82\times10^{-2}$ &  $1.89\times10^{-3}$ & $5.04\times10^{-3}$ & $1.58\times10^{-4}$ \\
 B-splines & $5.97\times10^{-2}$ & $1.88\times10^{-3}$  & $6.31\times10^{-3}$ & $1.99\times10^{-4}$ \\
\hline                                            
\end{tabular}
\end{table}

\begin{itemize}

\item \textbf{Influence of turbulence} The PSF feature that affects the interpolation methods the most is the presence of a Kolmogorov (KM) turbulence in ellipticity. Fig.~\ref{fig:turbulent_non-turbulent} illustrates how erratic the spatial variation pattern of ellipticity can become in the presence of KM turbulence. It is clear that a prediction algorithm faces a much more challenging task on turbulent images than on images with more regular PSF patterns.
To highlight this, we have averaged in Tables~\ref{table:Star Challenge statistics non-turbulent} and \ref{table:Star Challenge statistics turbulent} the metrics $E$ and $\sigma$ separately over turbulent and non-turbulent sets. Comparing these two tables shows that $E(e)\sim10^{-4}$ and $\sigma(e)\sim10^{-5}$ on non-turbulent sets, whereas $E(e)\sim10^{-2}$ and $\sigma(e)\sim10^{-3}$ on turbulent sets. This represents a $\sim$100-fold decrease in accuracy and error on the mean. This effect can also be seen on the plots of $E(e)$ in Figs.~\ref{fig:plots metrics part 1} for and \ref{fig:plots metrics part 2}.

We also observe that, on sets without a KM spectrum, all interpolators evaluated in this paper typically reach \mbox{$\sigma_{sys}^2\sim10^{-8}$} already beyond the \mbox{$\sim10^{-7}$} goal of next-generation space-based weak lensing surveys. In contrast, sets with turbulent PSF do not match that requirement, with \mbox{$\sigma_{sys}^2\sim10^{-6}$}.

\begin{figure*}[t]
\resizebox{\hsize}{!}{\includegraphics[trim=4mm 4mm 4mm 4mm, clip]{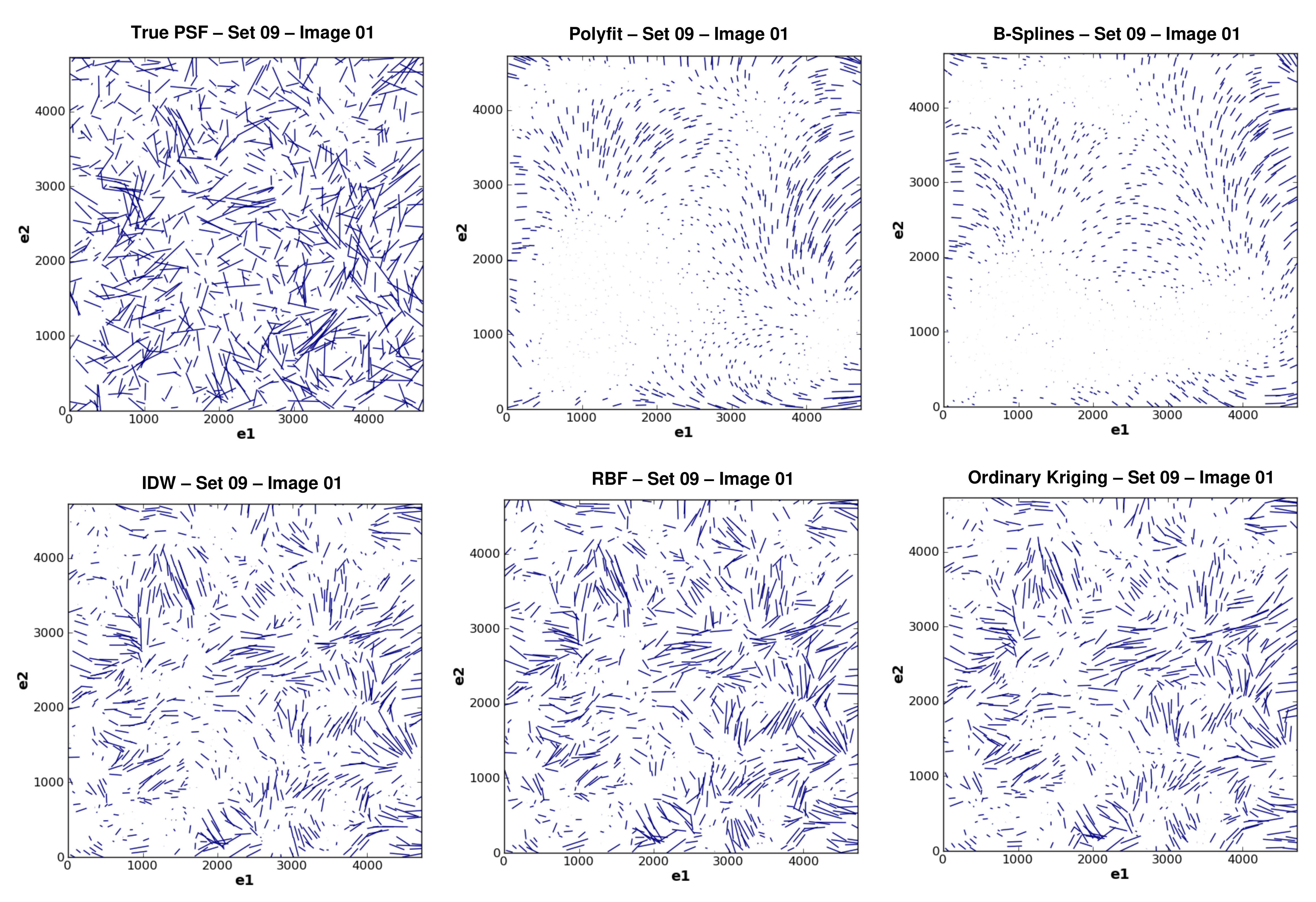}}
\caption{An illustration of how the various interpolation methods studied in this article handled a turbulent PSF, which in this case is the first image of set 9. The true ellipticities are plotted on the upper-left corner of the Figure and the remaining plots show the predictions of each methods. The largest whisker in the upper-left corner plot corresponds to an ellipticity of 0.38.}
\label{fig:interpolating ellipticities on a turbulent set}
\end{figure*}
%Before analyzing the performance of individual interpolation schemes in more detail, we can already make the following  \mbox{observations}:

The similarities between $E(e)$ and $\sigma(e)$ values for RBF, IDW and Kriging in Table~\ref{table:Star Challenge statistics turbulent} suggest these methods behave more or less the same when confronted with \mbox{turbulent} ellipticities. To check this, we have have compiled in Fig.~\ref{fig:interpolating ellipticities on a turbulent set} the true ellipticity pattern of turbulent set 9 along with the actual predictions of the same pattern by all five interpolators. The same metrics for Polyfit and B-splines show that these global methods are even more handicapped by the presence of a KM spectrum.

Turbulence also makes the spatial distribution of the FWHM less predictable and the methods are affected to various degrees: RBF, IDW, Polyfit and B-Spline are little influenced with similar $E(R^2)$ and $\sigma(R^2)$ values in Tables~\ref{table:Star Challenge statistics non-turbulent} and \ref{table:Star Challenge statistics turbulent} and on the corresponding plots in Figs.~\ref{fig:plots metrics part 1} and \ref{fig:plots metrics part 2}. The only one really impacted is Kriging. \\

\item \textbf{Influence of star density} Following the discussion of Sect.~\ref{subsection:alternative PSF interpolation schemes}, we expect the local interpolation methods to be more accurate than global ones on images with higher star density but see their performance degrade on sparser star fields. Such local interpolators base their predictions on observations found in local neighborhoods and should therefore be in position to take advantage of any additional available. On the other hand, they should suffer comparatively more from insufficient sampling when the data is too sparse. This is indeed what we observe in the IDW plot Fig.~\ref{fig:plots metrics part 1}, but the conclusion is less clear regarding RBF and Kriging: these schemes are indeed more accurate on denser sets when it comes to estimate the FWHM but the reverse is seen concerning ellipticities (plots Figs.~\ref{fig:plots metrics part 1} and \ref{fig:plots metrics part 2}). This is mostly noticeable on non-turbulent sets and may be caused by some overfitting taking place  on denser ellipticity fields. This phenomenon does not occur on FWHM possibly because the FWHM spatial distribution is generally smoother than that of ellipticities in the Star Challenge dataset.\\
We also expect the global interpolators B-splines and Polyfit to be little affected by difference in star density, since such schemes attempt to find a regression surface that takes all available data into account but at the same time minimize the overall bias through the least squares criterion. Such a surface tends to smooth out small-scale variations, mostly capturing broad features in the image. The corresponding predictions may become less accurate but, on the other hand, remain little influenced by sampling differences. This is exactly what we find in the plots of Polyfit and B-splines Fig.~\ref{fig:plots metrics part 2}. The smoothness of the prediction surfaces of Polyfit and B-splines compared to that of local interpolators is clearly noticeable in Fig.~\ref{fig:interpolating ellipticities on a turbulent set}.\\

\item \textbf{Influence of the PSF model and size, masking and telescope effects}  Although some interpolators do better than others on a particular PSF models, each individual scheme perform equally well on Moffat and Airy images. This can be seen, for example, on fiducial sets 1 and 8 where the error statistics on Moffat or Airy sets are almost identical for a given method. The same can be said of the influence of FWHM, masking and telescope effects. We also observe star size to have a negligible impact on $E(e)$ for all methods, but we clearly see that $E(R^2)$ significantly increases (resp. decreases) for star fields with smaller (resp. larger) FWHM. Finally, all methods reach a slightly higher accuracy on masked images, especially with 6-fold masks.

\end{itemize}

\subsection{Results from individual interpolation methods}
\label{subsection:results from individual interpolation methods}

\begin{itemize}

\item \textbf{Interpolation with radial basis functions (RBF)} As shown in our previous discussion, the RBF interpolation scheme is the overall winner of our evaluation. According to our benchmarks, ellipticity patterns were best estimated by a \emph{linear kernel} function, whereas a \emph{thin-plate kernel} was more effective on FWHM values. A neighborhood size between 30 and 40 stars was used. Refer to Sect.~\ref{subsection RBF interpolation} and Table~\ref{table:popular RBF kernels} for a description of RBF interpolation and the definitions of these kernels. That combination of linear and thin-plate kernels yields very competitive error statistics on both turbulent and non-turbulent sets: Tables~\ref{table:Star Challenge statistics non-turbulent} and \ref{table:Star Challenge statistics turbulent} as well as plots Fig.~\ref{fig:plots metrics part 1} show RBF is the most accurate on turbulent sets whereas its results on non-turbulent sets are the second best behind ordinary Kriging. The \mbox{possibility} of selecting the most suitable kernel for a given PSF patterns is a very attractive feature of RBF interpolation.\\

\item \textbf{Inverse distance weighted interpolation (IDW)} The IDW methods (see~Sect.~\ref{subsection:IDW}) obtains the second best average $E(e)$ behind RBF over all sets as seen in Table~\ref{table:new Star Challenge statistics}. It does so thanks to very competitive $E(e)$ results on turbulent sets, just behind RBF (Table~\ref{table:Star Challenge statistics turbulent}). But IDW's estimates of the FWHM on non-turbulent sets are by far the worst of all five interpolation algorithms, both on ellipticity and size (Table~\ref{table:Star Challenge statistics non-turbulent}). As found in Sect.~\ref{subsection:influence of PSF features simulated in the image}, IDW looks quite sensitive to variations in star density. In fact, we observe that IDW underperforms on star fields with low-density and smaller FWHM (sets 5, 6, 13, 14, 25, 26). We were unable to find a setup that significantly improves that level of accuracy, which suggests the method has difficulty coping with such constraints in density and size.\\
All in all, IDW performs quite well overall, knowing it is based on a very simple interpolation algorithm, with fewer adjustable parameters than RBF or ordinary Kriging (see Sect.~\ref{subsection:IDW}). 
%The IDW results were obtained by selecting a weighting power $\beta=2$ and a neighborhood size in the range $5\--15$ stars.
\\

\item \textbf{Interpolation with ordinary Kriging (OK)} Despite its reputation of best interpolator on spatially-scattered data, ordinary Kriging, introduced in Sect.~\ref{subsection:kriging}, arrives only third behind RBF and IDW when considering error statistics in Table~\ref{table:new Star Challenge statistics}. As see in shown in Table~\ref{table:Star Challenge statistics non-turbulent} and plots Fig.~\ref{fig:plots metrics part 2}, Kriging's estimates on non-perturbed sets are the best of all five methods. But this cannot compensate for its relatively poor performance on estimating the FWHM on turbulent sets, as shown in the value of $E(R^2)$ in Table~\ref{table:Star Challenge statistics turbulent}. The reason for this is probably related to the significant spatial drift of the FWHM values across the image. The condition of intrinsic stationarity required by ordinary Kriging is no longer fulfilled in some areas, especially near the edges of the image. As a result, we were forced to reduce the size of the search neighborhood over which the Kriging weights are calculated, which leads to a loss in accuracy in the corresponding regions. Kriging variants with ability to correct such a drift, like Universal Kriging, would probably achieve better results. Also, our implementation of Kriging for the Star Challenge assumes spatial isotropy, even though experimental variograms for ellipticity on non-turbulent sets also show evidence of geometric anisotropy, A more sophisticated implementation could have corrected these effects by rescaling and rotating coordinate axes along the direction of maximum spatial continuity.\\

\item \textbf{Polynomial fitting (Polyfit)} The results of Polyfit are of particular interest since polynomial fitting is currently the method of choice for modeling spatial variations of a PSF in lensing studies (see Sects.~\ref{section:existing PSF interpolation schemes} and \ref{subsection:polynomial fitting}). polynomial fitting performs relatively well on non-turbulent sets with $E(e)$ and $E(R^2)$ statistics fairly close to those of RBF (Table \ref{table:Star Challenge statistics non-turbulent}). However, the corresponding statistics on turbulent sets are significantly worse that those achieved by local methods, as seen in Table~\ref{table:Star Challenge statistics turbulent}. This confirms the conclusion of Sect.~\ref{subsection:polynomial fitting} whereby polynomials have difficulty coping with small or rapid variations found in a PSF pattern. Low-degree polynomials generally produce satisfactory result but tend to underfit the data, which leads to suboptimal accuracy. The resulting interpolation surfaces are characteristically smooth, as clearly observed in the Polyfit plot of Fig.~\ref{fig:interpolating ellipticities on a turbulent set}. The Star Challenge images without KM power spectrum are smooth enough for Polyfit to approach the accuracy of RBF and ordinary Kriging. These results were obtained with a fifth-degree polynomial, higher degrees degrading the fit.\\

\item \textbf{Interpolation with Basis splines (B-splines)} Polynomial splines are generally considered superior for interpolation than simple polynomials as explained in Sect.~\ref{subsection:spline interpolation}, and we would have expected B-splines to achieve better results than Polyfit on the Star Challenge data. But this is not reflected in the averaged results from Tables~\ref{table:new Star Challenge statistics}. The level of accuracy reached by both interpolators is nevertheless of the same order. \\ 
As seen in Table~\ref{table:Star Challenge statistics non-turbulent} and plots Fig.~\ref{fig:plots metrics part 2}, the ellipticity estimates from B-splines are superior to those of Polyfit on non-turbulent sets and of similar accuracy on turbulent ones. This tends to confirm the better ability of splines to capture small-scale and rapid variations in the data than polynomials. The results show, however, errors $E(R^2)$ on the FWHM much larger for B-splines than for Polyfit, which explains the relative lower performance compared to Polyfit. The FWHM spatial distribution being overall quite smooth in the Star Challenge images, this result suggests polynomials may be better suited than splines for modeling smoothly-varying patterns of variation. Combining both schemes may also be worth investigating.\\

%Tables~\ref{table:new Star Challenge statistics}, \ref{table:Star Challenge statistics non-turbulent} and \ref{table:Star Challenge statistics turbulent} and the corresponding plots in Fig.~\ref{fig:plots metrics part 2} summarize the results of Polyfit. On the other hand, the global

\end{itemize}

\section{Conclusions}
\label{section:conclusions}

%After a literature review of the use of PSF interpolation in weak lensing (Sect.~\ref{section:existing PSF interpolation schemes}), we criticized the commonly-used polynomial interpolation scheme and provided an overview of the abovementioned techniques in Sect.~\ref{section:PSF interpolation schemes}.

The GREAT10 Star Challenge gave us the opportunity to evaluate several interpolation methods on spatially-varying PSF fields:
\begin{itemize}
\item Two global, approximate and deterministic spatial interpolation schemes: polynomial fitting (Polyfit) and basis splines (B-splines).
\item Two local, exact and deterministic techniques relying on \mbox{inverse distance weighting} (IDW) and  \mbox{radial basis functions} (RBF).
\item An implementation of ordinary Kriging, a local, exact and stochastic spatial prediction method, frequently used in Geostatistics and environmental sciences.
\end{itemize}

We used a three-stage PSF estimation pipeline, which we described in Sect.~\ref{subsection:prediction pipeline} and Appendix~\ref{section:appendix PSF prediction pipeline}. Elliptical Moffat profiles were fitted to the stars contained in each Star Challenge image and then estimated and reconstructed at new positions in the same image using  one of the five interpolation schemes listed above.

That approach proved quite successful since it allowed us to win the GREAT10 Star Challenge.
We were, however, disappointed by the relatively high $\sigma_{sys}^2$ values reached, of the order of $10^{-4}$, i.e., still far from the $\sigma_{sys}^2 \lesssim10^{-7}$ target demanded by future large weak lensing surveys. The lack of accuracy could be traced to the suboptimal fitting of Airy PSF profiles by our pipeline and not to a deficiency in the PSF interpolation methods. However, this issue made it difficult to unambiguously conclude on the level of accuracy of individual interpolation algorithms, which is the main objective of this article.

In order to measure errors purely due to interpolation and only these, we used the true input ellipticity and FWHM catalog for the input Star Challenge images instead of our fitted estimates for these quantities. We also chose new metrics, better suited than the P-factor for assessing estimates on ellipticity and size. The results are summarized in Tables~\ref{table:new Star Challenge statistics}, \ref{table:Star Challenge statistics non-turbulent} and \ref{table:Star Challenge statistics turbulent} along with the corresponding plots in Figs.~\ref{fig:plots metrics part 1} and \ref{fig:plots metrics part 2}. 
We highlight our main \mbox{conclusions} below.

\begin{itemize}

\item Table~\ref{table:new Star Challenge statistics} shows the overall $E(e)$ and $E(R^2)$ errors to be of the order of $10^{-2}$ and $10^{-3}$ respectively. Fig~\ref{table:Star Challenge statistics non-turbulent}, however indicates that $E(e)\sim10^{-4}$ and $E(R^2)\sim10^{-3}$ on images devoid of Kolmogorov turbulence, to be compared with the $E(e)\lesssim10^{-3}$ and $E(R^2)\lesssim10^{-3}$ estimated requirements of future next-generation surveys. Although the Star Challenge PSF fields lack realism in certain aspects, this suggests that the best methods, RBF, IDW and OK, may already be suitable for space-based surveys where turbulence is absent.\\

\item All interpolation methods see their accuracy drastically degraded in images where atmospheric turbulence effects have been simulated, with $E(e)$ and $E(R^2)$ errors increased by a factor of $\sim100$. The better performance on turbulent images of RBF, IDW and OK compared to Polyfit and B-splines in the GREAT10 Star Challenge, suggests local methods may be able to better cope with turbulence than global ones. We note, however, that these results are only valid for the specific turbulence model used in the simulations and would have to be confirmed on real data.\\

\item After turbulence, the factors influencing results the most are the density of stars and their size. As far as density is concerned, local methods are more impacted than global ones and generally improve their estimates on denser sets much more than global methods. A similar conclusion is reached concerning local methods as far as PSF size is concerned. However, the results suggest both global and local methods have difficulty coping with objects smaller than the fiducial FWHM of $3$ pixels. Among all methods, IDW suffered the most from sparse star fields with small FWHM.\\

\item The RBF interpolator proved the most accurate, reaching the best results on both turbulent and non-turbulent sets. The use of kernel functions brings additional versatility compared to a simpler interpolator like IDW, while avoiding the complexity of Kriging. The selection of the most suitable kernel function and associated parameters can be greatly simplified by the use of cross-validation or Jackknifing. These techniques, as shown in Sect. \ref{subsection:cross-validation}, can prove very helpful to tune the run-time parameters of an interpolation schemes and evaluate the accuracy of its results.\\

\item Despite its simplicity, the IDW interpolation method obtained better than expected results, outperforming polynomials and splines in the simulations. Fast and easy to tune, it could potentially constitute a simple alternative/complement to polynomials before trying more elaborate interpolation schemes such as Kriging or RBF.\\

\item Ordinary Kriging is, in our opinion, potentially the most accurate method as shown especially by its results on non-turbulent images. However, the FWHM spatial distributions in the Star Challenge have a significant spatial drift that the standard ordinary Kriging algorithm is unable to correct. Another Kriging variant such as Universal Kriging would possibly have proved more accurate. It remains that Kriging, because of its sophistication, is more difficult and time consuming to operate than the other interpolators we evaluated.\\

\item  Overall, our analysis of the Star Challenge results suggests local \mbox{interpolators} should be preferred over global ones based on splines and polynomials. However, one should bear in mind that (1) these results are based on simulated data where star images are isolated, bright enough and well sampled; (2) the spatial variation of the PSF as simulated in GREAT10 may tend to favor local interpolators over global ones. We strongly believe, nevertheless, that local interpolation schemes for PSF interpolation have the potential to improve the accuracy of existing and future ground-based lensing surveys and deserve to be investigated further.

\end{itemize}

\begin{acknowledgements}
We thank Tom Kitching for his help and especially for \mbox{providing} us with the true input ellipticity and FWHM catalog of the Star Challenge PSF images. We also acknowledge support from the International Space Science Institute (ISSI) in Bern, where some of this research has been discussed. This work is supported by the Swiss National Science Foundation (SNSF). 
\end{acknowledgements}

\bibliographystyle{aa}
\bibliography{../Articles}

\begin{appendix}

\section{Our PSF prediction pipeline (PSFPP)}
\label{section:appendix PSF prediction pipeline}

\subsection{Overview}
\label{subsection:pipeline overview}

The PSF prediction pipeline used in the Star Challenge is outlined in Fig.~\ref{fig:PSF prediction pipeline}. A PSF field is fed into the pipeline and goes through three processing stages:
\begin{enumerate}
\item \emph{Fitting} stage: the Moffat PSF model described in Sect.~\ref{subsection:PSF model} is fitted to each star at known position $(x_c,y_c)$ in the PSF field image. A catalog is produced, containing a set of fitted parameters $\big\{(x_c,y_c); (e_1,e_2); \phi, (\alpha,\beta)\big\}$ for each star. Instead of an out-the-box minimizer, we employ a custom minimizer we developed at the EPFL Laboratory of astrophysics and well suited to fitting faint and noisy images like those frequently found in weak lensing. The minimizer uses an "adaptive cyclic coordinate descent algorithm" that find a local minimum with the lowest $\chi ^2$ of the residuals. That same minimizer has also been used in the version of the \emph{gfit} shear measurement method that competed in the GREAT10 Galaxy Challenge \citep{G10results}. The star images processed by the minimizer are $16\times16$-pixel cutouts instead of the original $48\times48$-pixel postage stamps.\\

\item \emph{Prediction} stage: 
\begin{itemize} 
\item First, an analysis of the spatial distribution of each parameter across the image is performed. In particular, all separation distances between stars are recorded in the form of \emph{ KD-trees} \citep{Bentley1975} for efficiently finding the nearest neighboring stars located within a given separation distance $\Arrowvert \mathbf{h} \Arrowvert$.
\item Second, a spatial prediction scheme is applied to estimate the values $Z_p'(x_i',y_i')$ of the parameter $p$ at asked locations $(x'_i,y'_i)$, given the fitted parameter values $Z{_p}(x_i,y_i)$ obtained in the previous stage. One of the four methods described in section \ref{section:PSF interpolation schemes} is applied here.\\ \end{itemize}
% Subsection \ref{subsection:cross-validation} outlines some techniques we used to evaluate and improve the accuracy of the predictions.\\

\item \emph{Reconstruction} stage: All stars in a PSF field are reconstructed based on the elliptical Moffat model described in Sect. \ref{subsection:PSF model}, but using the parameters predicted for that star during the \emph{Prediction} stage.
\end{enumerate}

\subsection{Pipeline implementation and configuration}
\label{subsection:pipeline implementation}
The pipeline code is written in Python, a programming language known for its power, flexibility and short development cycle. The usual standard Python libraries are used, notably: NumPy, SciPy, PyFITS and matplotlib. SciPy is the standard scientific library for Python. Most of its functions are thin Python wrappers on top of fortran, C and C++ functions. SciPy takes advantage of installed optimized libraries such as LAPACK (Linear Algebra PACKage) library \citep{Anderson1990}. We employ the cross-validation and Jackknifing resampling techniques (see \ref{subsection:cross-validation}) to tune the run-time parameters for the interpolation schemes and evaluate the accuracy of the results. We highlight below a few aspects related to the implementation of the methods. 
%We now describe in more details the implementation of the interpolation methods we .\newline

%The Scientific Python library SciPy provided the implementation for radial basis functions (RBF) and spline interpolation. \newline\indent
%\renewcommand{\labelitemi}{$\diamond$}
\begin{itemize} 
\item IDW: the code for Inverse Distance Weighted interpolation is written in Python, based on Eq.~(\ref{eq:generic interpolation formula}) with weighting factors specified by (\ref{eq:IDW weights}). The free parameters are the power factor $\beta$ and the neighborhood size $N$ (see \ref{subsection:IDW}). A configuration with $\beta=2$, with $5 \leq N \leq 15$ depending on the density of stars in images gives the best results according to our tests. \\

\item RBF: we use the \texttt{rbf()} interpolation function available in the \emph{SciPy} \emph{interpolate} module. The number of parameters to tune is greater compared to IDW: a kernel function chosen among those listed in Table~\ref{table:popular RBF kernels}; the neighborhood search size $N$; a shape parameter $\epsilon$ for the \emph{multiquadric}, \emph{inverse multiquadric} and \emph{Gaussian} kernels; and a last parameter for controlling the smoothness of the interpolation (see \ref{subsection RBF interpolation}). Only the \emph{linear }, \emph{thin-plate} and \emph{multiquadric} kernels gave stable enough predictions. Choosing $25 \leq N \leq 30$ and disabling smoothing (i.e. use exact interpolation) yielded the best cross-validation and Jackknifing results for the chosen kernels.\\

\item splines: we have selected the \texttt{bisplrep()} and \texttt{bisplev()} bivariate B-spline interpolation functions provided by the SciPy \emph{interpolate} module. These functions are Python wrappers on top of the fortran FITPACK package \citep{Dierckx1995}. The underlying algorithms follow the \emph{constructive} approach for spline interpolation described in \ref{subsection:spline interpolation} and are specified in \citet{Dierckx1980}. The main parameters affecting the interpolation are the degree $p$ of the spline, the number of knots $N$ and a smoothing factor $s$. We have fixed $p$ to $3$ but let the algorithm automatically set $N$ and $s$. \\

\item Kriging: we have used our own custom-developed Python code of ordinary Kriging (see~Sect.~\ref{subsection:kriging}). The Kriging used in the Star Challenge and in this article is isotropic and does not implement any spatial anisotropy or drift correction scheme. 

%\item Kriging: we have used our own custom-developed Python code of ordinary Kriging (see~Sect.~\ref{subsection:kriging}). The version used in the Star Challenge and in this article does not implement any spatial anisotropy or drift correction scheme. The variograms for ellipticity on non-turbulent sets show evidence of geometric anisotropy, that could have been corrected by rescaling and rotation of the axes. Turbulent sets

 %The shapes of the variograms on non-turbulent sets show evidence of geometric anisotropy, that could have been corrected by rescaling and rotation of the axes. Turbulent setÉ 

The accuracy of the ordinary Kriging interpolation scheme was influenced by the following set of parameters: 
\begin{itemize} 
\item The interpolation range, i.e. the range in pixels used for interpolation. Depending on the images, we chose a circular area with a radius between $700$ and $1000$ pixels from the center of the $4800\times4800$ PSF field.
\item Lag distance $h$ in pixels. We used values in the range $100 \leq h \leq 300$ depending on the image and the PSF model parameter to estimate.
\item The number of observations $N$ in Eq.~(\ref{eq:generic interpolation formula 2}) to include in the neighborhood: we used $5 \leq N \leq 20$ depending on the image star density. 
\item Tolerance distance $\Delta h$ (pixels) and angle $\Delta \theta$ considered when locating neighboring observations. As a rule of thumb, we selected $\Delta h \thickapprox h/2$ and $\Delta \theta = 22.5 ^\circ$.
\item A theoretical variogram model such as those listed in Table~\ref{table:variogram models}. The \mbox{experimental variograms} were fitted using the \emph{Levenberg-Marquardt} least-squares \texttt{leastsq} routine from the SciPy \emph{optimize} \mbox{module}. The program dynamically selected the theoretical variogram models and parameters that produced the best fit.\\
\end{itemize} 

\item  The Polyfit code is based on the \texttt{leastsq()} function from the \emph{SciPy} \emph{optimize} Python module. A least-squares fit to a bivariate polynomial of degree 5 gave the best estimates.
\end{itemize} 

\end{appendix}

%\begin{appendix}

%\section{Performance metrics per image sets}
%\label{section:appendix performance metrics per set}

%\end{appendix}

\end{document}